\documentclass[10pt,twocolumn,letterpaper]{article}

\usepackage{cvpr}
\usepackage{pgfplots}
\usepackage{pgf,tikz}
\usepackage{times}
\usepackage{epsfig}
\usepackage{graphicx}
\usepackage{amsmath}
\usepackage{amssymb}
\usepackage{overpic}
\usepackage{bm}
\usepackage{wrapfig}

\long\def\epigram#1{% We use \long\def to allow the argument
    \bigskip       % extra space above.
    \setbox0=\hbox{#1} % put the text into a box
    \ifdim\wd0>\critwidth %check the length
         \centerline{\vtop{\hsize=\episize\noindent#1}}
    \else          % otherwise, we just center the text.
         \centerline{\box0}
    \fi
    \smallskip     % extra space for the end.
}
     
\newdimen\critwidth          \critwidth=.85\hsize
\newdimen\episize            \episize=.8\hsize

\newcommand{\X}{\mathcal{X}}
\newcommand{\Y}{\mathcal{Y}}

\usepackage[breaklinks=true,bookmarks=false]{hyperref}

\cvprfinalcopy % *** Uncomment this line for the final submission

\def\cvprPaperID{3421} % *** Enter the CVPR Paper ID here

\ifcvprfinal\fi
\begin{document}

%%%%%%%%% TITLE
\title{Isospectralization, or how to hear shape, style, and correspondence}

\author{
Luca Cosmo\\University of Venice\\{\tt\small luca.cosmo@unive.it}
\and
Mikhail Panine\\\'Ecole Polytechnique\\{\tt\small mpanine@lix.polytechnique.fr}
\and
Arianna Rampini\\Sapienza University of Rome\\{\tt\small rampini@di.uniroma1.it}
\and
\hspace{-2mm}Maks Ovsjanikov\hspace{-2mm}\\\'Ecole Polytechnique\\{\tt\small maks@lix.polytechnique.fr}
\and
Michael M. Bronstein\\\hspace{-5mm}Imperial College London / USI\hspace{-5mm}\\{\tt\small m.bronstein@imperial.ac.uk}
\and
Emanuele Rodol\`a\\Sapienza University of Rome\\{\tt\small emanuele.rodola@uniroma1.it}
}

\maketitle
%\thispagestyle{empty}

%%%%%%%%% ABSTRACT
\begin{abstract}
The question whether one can recover the shape of a geometric object from its Laplacian spectrum (`hear the shape of the drum') is a classical problem in spectral geometry with a broad range of implications and applications. 
While theoretically the answer to this question is negative (there exist examples of iso-spectral but non-isometric manifolds), little is known about the practical possibility of using the spectrum for shape reconstruction and optimization. 
In this paper, we introduce a numerical procedure called {\em isospectralization}, consisting of deforming one shape to make its Laplacian spectrum match that of another. We implement the isospectralization procedure using modern differentiable programming techniques and exemplify its applications in some of the classical and notoriously hard problems in geometry processing, computer vision, and graphics such as shape reconstruction, pose and style transfer, and dense deformable correspondence. 
\end{abstract}

\section{Introduction}\label{sec:intro}
\vspace{-0.5ex}
Can one hear the shape of the drum? This classical question in spectral geometry, made famous by Mark Kac's eponymous paper~\cite{kac.drum}, inquires about the possibility of recovering the structure of a geometric object from its Laplacian
spectrum. Empirically, the relation between shape and its acoustic properties has long been known and can be traced back
at least to medieval bellfounders. However, while it is known that the spectrum carries many geometric and topological
properties of the shape such as the area, total curvature, number of connected components, etc., it is now known that
one cannot `hear' the metric.
Examples of high-dimensional manifolds that are isospectral but not isometric have been constructed in 1964
\cite{milnor.tori} (predating Kac's paper), but it took until 1992 to produce a counter-example of 2D polygons giving a
negative answer to Kac's question~\cite{cannothear,gordon1992isospectral}.

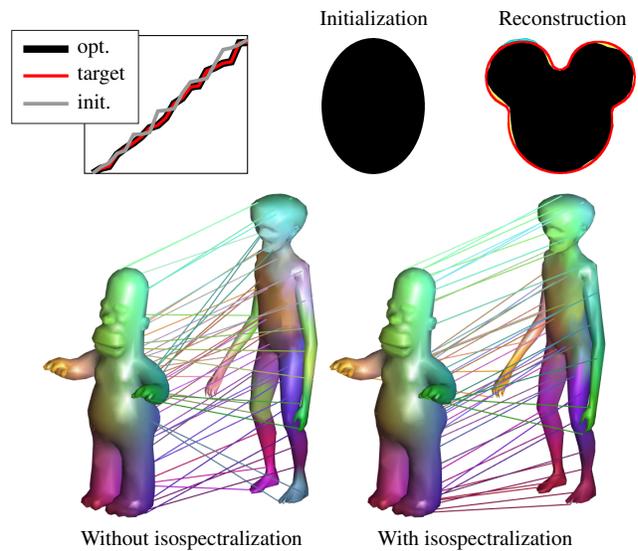
\begin{figure}[t!]
  \centering
\definecolor{mycolor1}{rgb}{0.6,0.6,0.6}%
\definecolor{mycolor2}{rgb}{0,0,0}%
\definecolor{mycolor3}{rgb}{1,0,0}%
\begin{tikzpicture}

\begin{axis}[%
width=0.26\linewidth,
height=0.22\linewidth,
at={(0.758in,0.481in)},
scale only axis,
xmin=0,
xmax=20,
ymin=0,
ymax=180,
xtick={},
xticklabels={},
ytick={},
yticklabels={},
ticks=none,
axis background/.style={fill=white},
legend style={legend cell align=left, align=left, draw=white!15!black},
legend style={at={(-0.45,0.7)},anchor=west}
]

\addplot [color=mycolor2, line width=2.8pt]
  table[row sep=crcr]{%
1	2.4208e-15\\
2	6.5446\\
3	9.1978\\
4	24.83\\
5	32.867\\
6	39.671\\
7	48.048\\
8	59.482\\
9	62.995\\
10	72.67\\
11	91.266\\
12	98.269\\
13	112.82\\
14	114.36\\
15	133.97\\
16	138.23\\
17	143.85\\
18	145.86\\
19	170.48\\
20	171.11\\
};
\addlegendentry{\footnotesize opt.}

\addplot [color=mycolor3, line width=1.2pt]
  table[row sep=crcr]{%
1	3.6816e-14\\
2	6.4937\\
3	9.1311\\
4	24.794\\
5	32.807\\
6	39.682\\
7	47.949\\
8	59.441\\
9	62.918\\
10	72.707\\
11	91.237\\
12	98.417\\
13	112.75\\
14	114.36\\
15	134\\
16	138.27\\
17	143.93\\
18	145.77\\
19	170.46\\
20	171.2\\
};
\addlegendentry{\footnotesize target}

\addplot [color=mycolor1, line width=1.4pt]
  table[row sep=crcr]{%
1	5.4018e-14\\
2	8.2265\\
3	13.581\\
4	26.081\\
5	29.721\\
6	50.849\\
7	51.19\\
8	53.577\\
9	81.288\\
10	82.604\\
11	84.619\\
12	104.94\\
13	119.41\\
14	121.54\\
15	123.99\\
16	143.25\\
17	161.68\\
18	163.89\\
19	168.58\\
20	174.78\\
};
\addlegendentry{\footnotesize init.}

\end{axis}
\end{tikzpicture}%
\hspace{0.8cm}
\begin{overpic}
[trim=0cm 0cm 0cm 0cm,clip,width=0.5\linewidth]{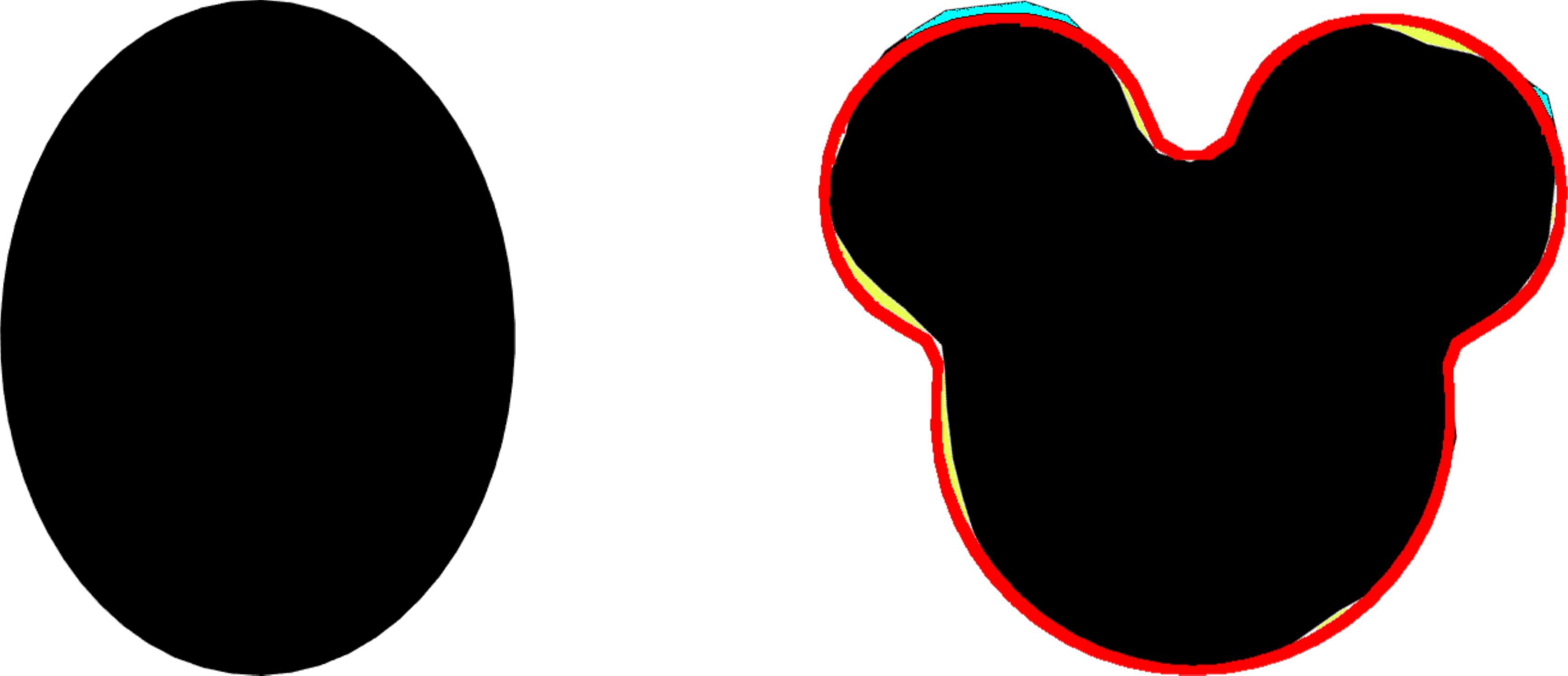}
%\put(-22,19){\footnotesize $\lambda$}
\put(-0.5,47.5){\footnotesize Initialization}
\put(56.5,47.5){\footnotesize Reconstruction}
\end{overpic}
\begin{overpic}
[trim=-0.4cm 0cm 0cm -1.1cm,clip,width=0.45\linewidth]{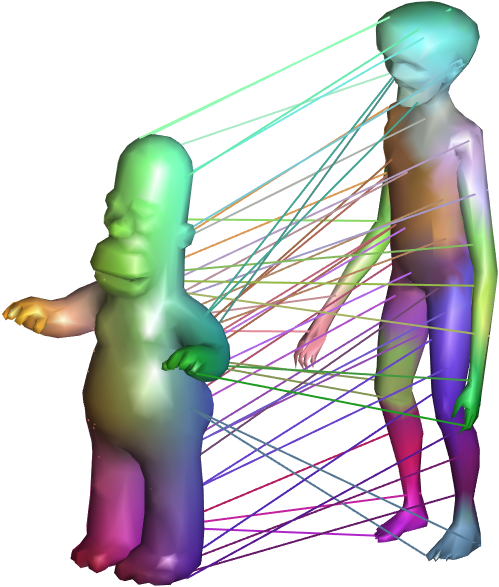}
\put(13,-8){\footnotesize{Without isospectralization}}
\end{overpic}
\begin{overpic}
[trim=-0.4cm 0cm 0cm -1.1cm,clip,width=0.45\linewidth]{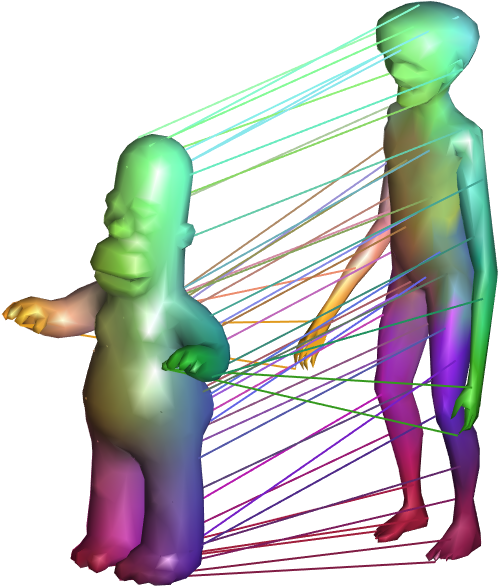}
\put(16,-8){\footnotesize{With isospectralization}}
\end{overpic}
\vspace{5mm}
  \caption{\label{fig:teaser}{\em Top row}: Mickey-from-spectrum: we recover the shape of Mickey Mouse from its first 20 Laplacian eigenvalues (shown in red in the leftmost plot) by deforming an initial ellipsoid shape; the ground-truth target embedding is shown as a red outline on top of our reconstruction. {\em Bottom row}: Aligning the Laplacian eigenvalues (`isospectralization') can be used as a preconditioning step for non-isometric deformable shape matching. We show the correspondence obtained with a baseline matching algorithm before (left) and after (right) isospectralization. Corresponding points are depicted with same color.}
\end{figure}

Nevertheless, the question of relation between shape and spectrum is far from being closed, from both theoretical and
practical perspectives.  Specifically, it is not yet certain whether the counterexamples are the rule or the exception.
So far, everything points towards the latter. In fact, there are known classes of manifolds in which the spectral
reconstruction is generically possible. See
\cite{zelditch1998revolution,zelditch2000spectral,zelditch2009inverse,hezari2010inverse} for such results.  Thus, it is
plausible that the theoretical existence of rather `exotic' counter-examples of non-isometric isospectral manifolds does not preclude the possibility of reconstructing the shape from its spectrum {\em in practice}.

This is exactly the direction explored in our paper.  We introduce a numerical procedure we call {\em
  isospectralization}, which consists in deforming a mesh in order to align its (finite) Laplacian spectrum with 
a given one.
We implement this procedure using modern differentiable programming tools used in deep learning applications and show
its usefulness in some of the fundamental problems in geometry processing, computer vision, and graphics.

For example, we argue that isospectralization (with some additional priors such as smoothness and enclosed volume) can in some cases be used to recover the structure of an object from its spectrum, thus practically hearing the shape of the drum (see
Figure~\ref{fig:teaser}, top row).

Outside of the rare counterexamples, the reconstruction is ambiguous up to intrinsic isometry. This ambiguity manifests
itself as a choice of an embedding of the mesh into $\mathbb{R}^3$.  This enables us to use the isospectralization
procedure to transfer style and pose across objects similarly to \cite{boscaini2015shape}: we initialize with a source
shape and apply isospectralization to obtain the eigenvalues of the target shape; the result is a shape in the pose of
the source shape but with geometric details of the target.

%Since the reconstruction is usually ambiguous up to isometry, the same procedure can be used to transfer style and pose across
%objects similarly to \cite{boscaini2015shape}: we initialize with a source shape and apply to isospectralization to obtain the eigenvalues of the target shape; the result is a shape in the pose the source shape but with geometric details of the target.  %\mikhail{This sentence is a bit unclear. We should paraphrase it to make it more precise}
%
Even more remarkably, we show that pre-warping non-isometric shapes by means of isospectralization can significantly
help in solving the problem of finding intrinsic correspondences between them (Figure~\ref{fig:teaser}, bottom row),
suggesting that our procedure could be a universal pre-processing technique for general correspondence pipelines.

\vspace{1ex}\noindent\textbf{Contribution.}
We consider the shape-from-eigenvalues problem and investigate its relevance in a selection of problems from computer vision and graphics. 
Our key contributions can be summarized as follows:
\begin{itemize}
\item Despite being highly non-linear and hard to compute, we show for the first time that the inverse mapping between a geometric domain and its Laplacian spectrum is addressable with modern numerical tools;

\item We propose the adoption of simple regularizers to drive the spectrum alignment toward numerically optimal solutions;

\item We showcase our method in the 2D and 3D settings, and show applications of style transfer and dense mapping of non-isometric deformable shapes.
\end{itemize}

\section{Related work}
The possibility of reconstructing shape from spectrum is of interest in theoretical physics \cite{kempf2010spacetime}, and has been explored by theoreticians since the '60s starting from Leon Green's question if a Riemannian manifold is fully determined by its (complete) spectrum~\cite{berger2012panoramic}. The isospectrality vs isometry question received a negative answer in the seminal work of Milnor~\cite{milnor.tori}, and additional counterexamples were provided by Kac~\cite{kac.drum} and Gordon~\etal~\cite{cannothear} to name some classical examples.  A complete survey of the theoretical literature on the topic is out of the scope of this paper; below, we only consider the far less well-explored {\em practical} question of how to realize metric embeddings from the sole knowledge of the (finite) Laplacian eigenvalues.

A related but more general class of problems takes the somewhat misleading name of {\em inverse eigenvalue problems}~\cite{chu2005inverse}, dealing with the reconstruction of a generic physical system from prescribed spectral data. Different formulations of the problem exist depending on the matrix representation of the system; in the majority of cases, however, at least partial knowledge of the {\em eigenvectors} is also assumed.

In the fields of computer vision and geometry processing, Reuter~\etal~\cite{reuter05,reuter06} investigated the informativeness of the Laplacian eigenvalues for the task of 3D shape retrieval. The authors proposed to employ the Laplacian spectrum as a global shape signature (dubbed the `shape DNA'), demonstrating good accuracy in distinguishing different shape classes. However, measuring the extent to which eigenvalues carry geometric and topological information about the shape was left as an open question.

More recently, there have been attempts at reconstructing 3D shapes from a full Laplacian matrix or other intrinsic operators~\cite{boscaini2015shape,corman2017functional}. Such methods differ from our approach in that they leverage the {\em complete} information encoded in the input operator matrix, while we only assume to be given the operator's eigenvalues as input. Further, these approaches follow a two-step optimization process, in which the Riemannian metric (edge lengths in the discrete case) is first reconstructed from the input matrix, and an embedding is obtained in a second step. As we will show, we operate ``end-to-end'' by solving directly for the final embedding.
It is worthwhile to mention that the problem of reconstructing a shape from its metric is considered a challenging problem in itself \cite{borrelli2012flat,chern18}. In computer graphics, several shape modeling pipelines involve solving for an embedding under a {\em known} mesh connectivity and additional extrinsic information in the form of user-provided positional landmarks~\cite{sorkine2004least}.

More closely related to our problem is the shape registration method of~\cite{hamidian16,hu17}. The authors propose to solve for a conformal rescaling of the metric of two given surfaces, so that the resulting eigenvalues align well. While this approach shares with ours the overall objective of aligning spectra, the underlying assumption is for the Laplacian matrices and geometric embeddings to be given. A similar approach was recently proposed in the conformal pre-warping technique of~\cite{schonsheck2018nonisometric} for shape correspondence using functional maps.

A related, but different, inverse spectral problem has been tackled in \cite{bharaj2015computational}. There, the task is to optimize the shape of metallophone keys to produce a desired sound when struck in a specific place. Prescribing the sound consists of prescribing a sparse selection of frequencies (eigenvalues) and the amplitudes to which the frequencies are excited when the key is struck. It is also desirable that the other frequencies be suppressed. This is different from the reconstruction pursued in our work, % In particular, in \cite{bharaj2015computational} the desired eigenvalues are not prescribed a particular position in the spectrum, they are merely required to be present somewhere in the spectrum; instead,
 since we prescribe a precise sequence of eigenvalues. Further, amplitude suppression in \cite{bharaj2015computational} is implemented by designing the nodal sets of specific {\em eigenfunctions}, thus bringing this type of approach closer to a partially described inverse eigenvalue problem~\cite[Chapter 5]{chu2005inverse}.

Perhaps most closely related to our approach are methods that have explored the possibility of reconstructing shapes from their spectrum in the case of coarsely triangulated surfaces~\cite{aasen2013shape} and planar domains \cite{panine2016towards}. These works also indicate that non-isometric isospectral shapes are exceedingly rare. Compared to the present paper, \cite{aasen2013shape} and \cite{panine2016towards} study shapes with a low number of degrees of freedom. There, the shapes are prescribed by fewer than 30 parameters, while we allow every vertex in the mesh to move.

\section{Background}
\vspace{1ex}\noindent\textbf{Manifolds.}
We model a shape as a compact connected 2-dimensional Riemannian manifold $\X$ (possibly with boundary $\partial\X$) embedded either in $\mathbb{R}^2$ (flat shape) or $\mathbb{R}^3$ (surface).
The intrinsic gradient $\nabla$ and the positive semi-definite Laplace-Beltrami operator $\Delta$ on $\X$ generalize the corresponding notions of gradient and Laplacian from Euclidean spaces to manifolds. In particular, $\Delta$ admits a spectral decomposition
 \begin{eqnarray}
\Delta \varphi_i(x) = \lambda_i \varphi_i(x)  & \,\,\,\,\,& x \in \mathrm{int}(\X) \\
\langle \nabla \varphi_i(x) , \hat{n}(x) \rangle = 0 &\,\,\,\,\,& x \in \partial\X\,, \label{eq:neu}
\end{eqnarray}
with homogeneous Neumann boundary conditions~\eqref{eq:neu}; here $\hat{n}$ denotes the normal vector to the boundary. 

\vspace{1ex}\noindent\textbf{Spectrum.}
The {\em spectrum} of $\X$ is the sequence of eigenvalues of its Laplacian. These form a discrete
set, which is a canonically ordered non-decreasing sequence:
\begin{equation}
0 = \lambda_1 < \lambda_2 \le \cdots\,,
\end{equation}
where $\lambda_1$ has multiplicity 1 due to the connectedness of $\X$; for $i>1$, the multiplicity of $\lambda_i$ is related to the intrinsic symmetries of $\X$.
The growth rate of the ordered sequence $(\lambda_i)$ is further related to the total surface area of $\X$ via Weyl's asymptotic law~\cite{weyl}:
\begin{equation}\label{eq:weyl}
\lambda_i \sim \frac{4\pi}{\int_\X dx} i\,,\quad i\to\infty\,.
\end{equation}

This result makes it clear that size can be directly deduced from the spectrum (\ie, one can ``hear the size'' of the drum). This fact can be used in a reconstruction algorithm, for example, by providing an initial embedding having approximately the sought area.

%Other geometric and topological properties such as the Euler characteristic~\cite[Theorem 3]{reuter06} can also be estimated from $(\lambda_i)$.

%eigenvalues encode topological and geometric properties of the manifold \cite{reuter06}

%\vspace{2ex}\noindent\textbf{Spectral geometry processing} is concerned with studying ... . Neumann boundary conditions (btw: why?).
%

%\vspace{2ex}\noindent\textbf{Isometry invariance. }
%
%Laplacian is invariant to isometries, and so are its eigenvalues/eigenspaces; in 2d only rigid, in 3d a lot of variability

\begin{figure}[t]
  \centering
\begin{overpic}
[trim=0cm 0cm 0cm 0cm,clip,width=0.85\linewidth]{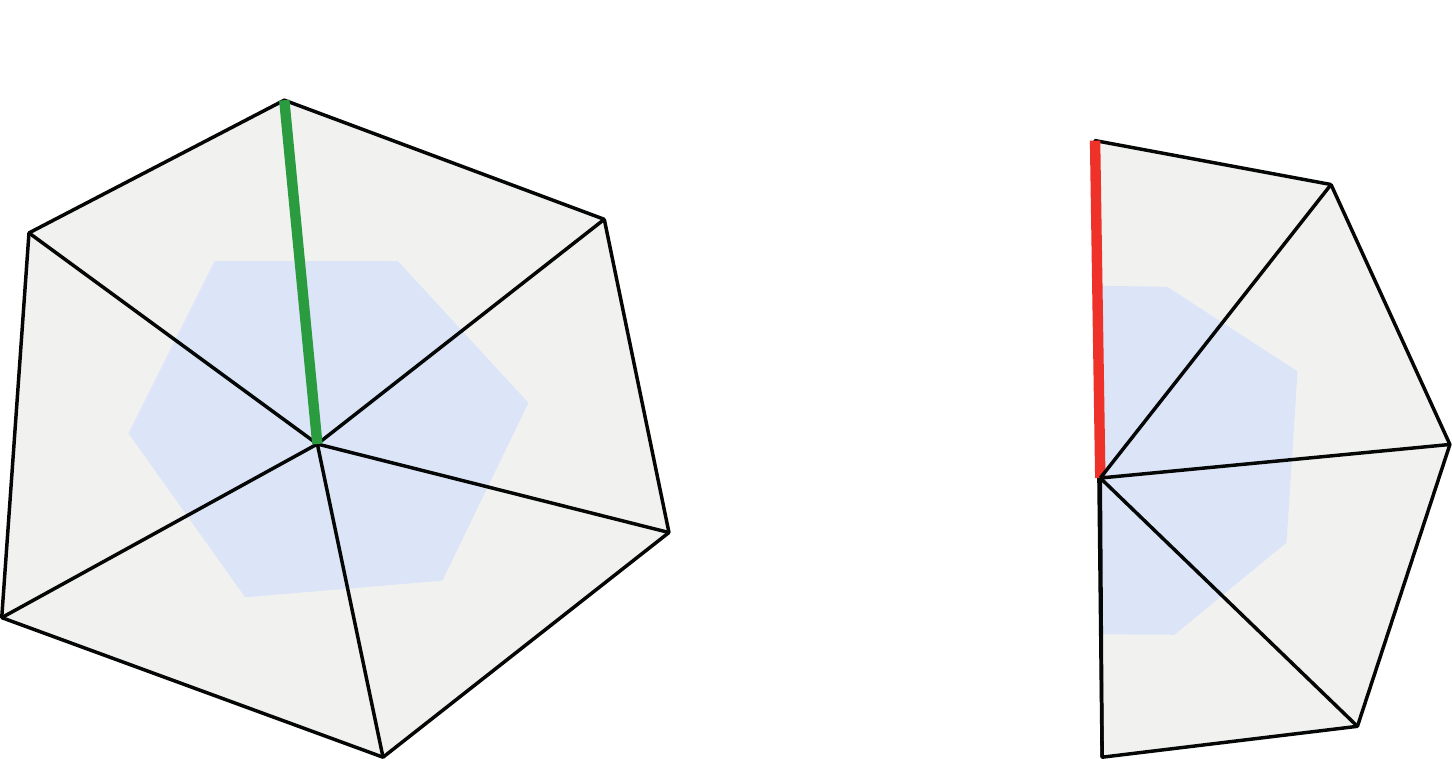}
\put(19,46){\footnotesize $\mathbf{v}^i$}
\put(20,17){\footnotesize $\mathbf{v}^j$}
\put(-2,36){\footnotesize $\mathbf{v}^k$}
\put(42.5,37){\footnotesize $\mathbf{v}^h$}
\put(74,44){\footnotesize $\mathbf{v}^i$}
\put(71,18){\footnotesize $\mathbf{v}^j$}
\put(92,39){\footnotesize $\mathbf{v}^h$}
\put(21.5,33.5){\footnotesize $\ell_{ij}$}
\put(8,26){\footnotesize $\ell_{jk}$}
\put(7,42.5){\footnotesize $\ell_{ki}$}
\put(30,42.5){\footnotesize $\ell_{hi}$}
\put(32,26.5){\footnotesize $\ell_{jh}$}
\put(70,30){\footnotesize $\ell_{ij}$}
\put(82,42.5){\footnotesize $\ell_{hi}$}
\put(85,28.5){\footnotesize $\ell_{jh}$}
\end{overpic}
\vspace{0.1cm}
  \caption{\label{fig:notation}Notation used in this paper. Edge $e_{ij}$ has length $\ell_{ij}$; triangle $F_{ijk}$ has area $A_{ijk}$. The shaded polygon denotes the local area element $a_j$ at vertex $v_j$.}
\end{figure}
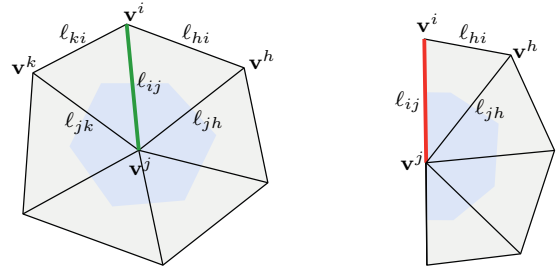

\vspace{2ex}\noindent\textbf{Discretization.}
In the discrete setting, our shapes $\X$ are approximated by manifold triangle meshes $X = (V,E,F)$ sampled at vertices $V=\{v_1,\dots,v_n\}$, and where each edge $e_{ij}\in E_i \cup E_b$ belongs to at most two triangle faces $F_{ijk}$ and $F_{jih}$. We denote by $E_i$ and $E_b$ the interior and boundary edges, respectively.
The discrete Riemannian metric is defined by assigning a length $\ell_{ij}>0$ to each edge $e_{ij}\in E$; see Figure~\ref{fig:notation} for the notation.

A $d$-dimensional {\em embedding} for $X$ is realized by assigning coordinates in $\mathbb{R}^d$ to the vertices $V$; these are encoded in a $n\times d$ matrix $\mathbf{V}$ containing $d$-dimensional vertex coordinates $\mathbf{v}^i$ for $i=1,\dots,n$ as its rows. Edge lengths can thus be written in terms of $\mathbf{V}$ as:
\begin{equation}
\ell_{ij}(\mathbf{V}) = \| \mathbf{v}^i - \mathbf{v}^j \|_2
\end{equation}
for all $e_{ij}\in E$.

The discrete Laplace-Beltrami operator assumes the form of a $n\times n$ matrix $\bm{\Delta} = \mathbf{A}^{-1} \mathbf{W}$, where $\mathbf{A}$ is a diagonal matrix of local area elements $a_i=\frac{1}{3}\sum_{jk:ijk\in F} A_{ijk}$, and $\mathbf{W}$ is a symmetric matrix of edge-wise weights,  defined in terms of the discrete metric as\footnote{It can be easily shown that this discretization is equivalent to the classical cotangent formulas~\cite{meyer2003discrete}, see e.g. \cite{jacobson2012cotangent}.}:
\begin{equation}\label{eq:w}
w_{ij}= \left\{ \begin{array}{ll}
         \frac{\ell^2_{ij}-\ell^2_{jk}-\ell^2_{ki}}{8 A_{ijk}} + \frac{\ell^2_{ij}-\ell^2_{jh}-\ell^2_{hi}}{8 A_{ijh}} & \mbox{if $e_{ij}\in E_i$}\\
        \frac{\ell^2_{ij}-\ell^2_{jh}-\ell^2_{hi}}{8 A_{ijh}} & \mbox{if $e_{ij} \in E_b$}\\
        -\sum_{k\neq i} w_{ik}  &  \mbox{if $i=j$}\end{array} \right.
\end{equation}

This discretization clearly depends on the mesh connectivity (encoded by edges $E$ and triangles $F$) and on the vertex coordinates $\mathbf{V}$ (via the lengths $\ell_{ij}$); since both play important roles in our reconstruction problem, we make this dependency explicit by writing $\bm{\Delta}_X(\mathbf{V})$.% instead of simply $\bm{\Delta}$.

%note: this definition implies Neumann boundary conditions

%we use the LB definition in terms of edge lengths; we don't do generalized eig, instead we normalize L; define a mesh: we optimize over vertex coordinates, but the optimization problem is aware of connectivity and actually uses it explicitly in a regularizer for regular tessellations
\section{Isospectralization}
Our approach builds upon the assumption that knowledge of a limited portion of the spectrum is enough to fix the shape of the domain, given some minimal amount of additional information which we phrase as simple regularizers.
We consider inverse problems of this form:
\begin{align}\label{eq:p1}
\min_{\mathbf{V}\in\mathbb{R}^{n\times d}} \| \bm{\lambda}(\bm{\Delta}_X(\mathbf{V})) - \bm{\mu} \|_\omega +  \rho_X(\mathbf{V})\,,
\end{align}
where $\mathbf{V}$ is the (unknown) embedding of the mesh vertices in $\mathbb{R}^d$, $\bm{\Delta}_X(\mathbf{V})$ is the associated discrete Laplacian, $\|\cdot\|_\omega$ is a weighted norm defined below, and $\bm{\mu},\bm{\lambda}\in\mathbb{R}_+^k$ respectively denote the input sequence and the first $k$ eigenvalues of $\bm{\Delta}_X(\mathbf{V})$. 
Function $\rho_X$ is a regularizer for the embedding, implementing the natural expectation that the sought solution should satisfy certain desirable properties. %The objective of an inverse eigenvalue problem is to construct a physical system that maintains a certain specific structure as well as that given spectral property.

\begin{figure}[t!]
  \centering
\vspace{0.1cm}
\begin{overpic}
[trim=0cm -1.7cm 0cm 0cm,clip,width=0.95\linewidth]{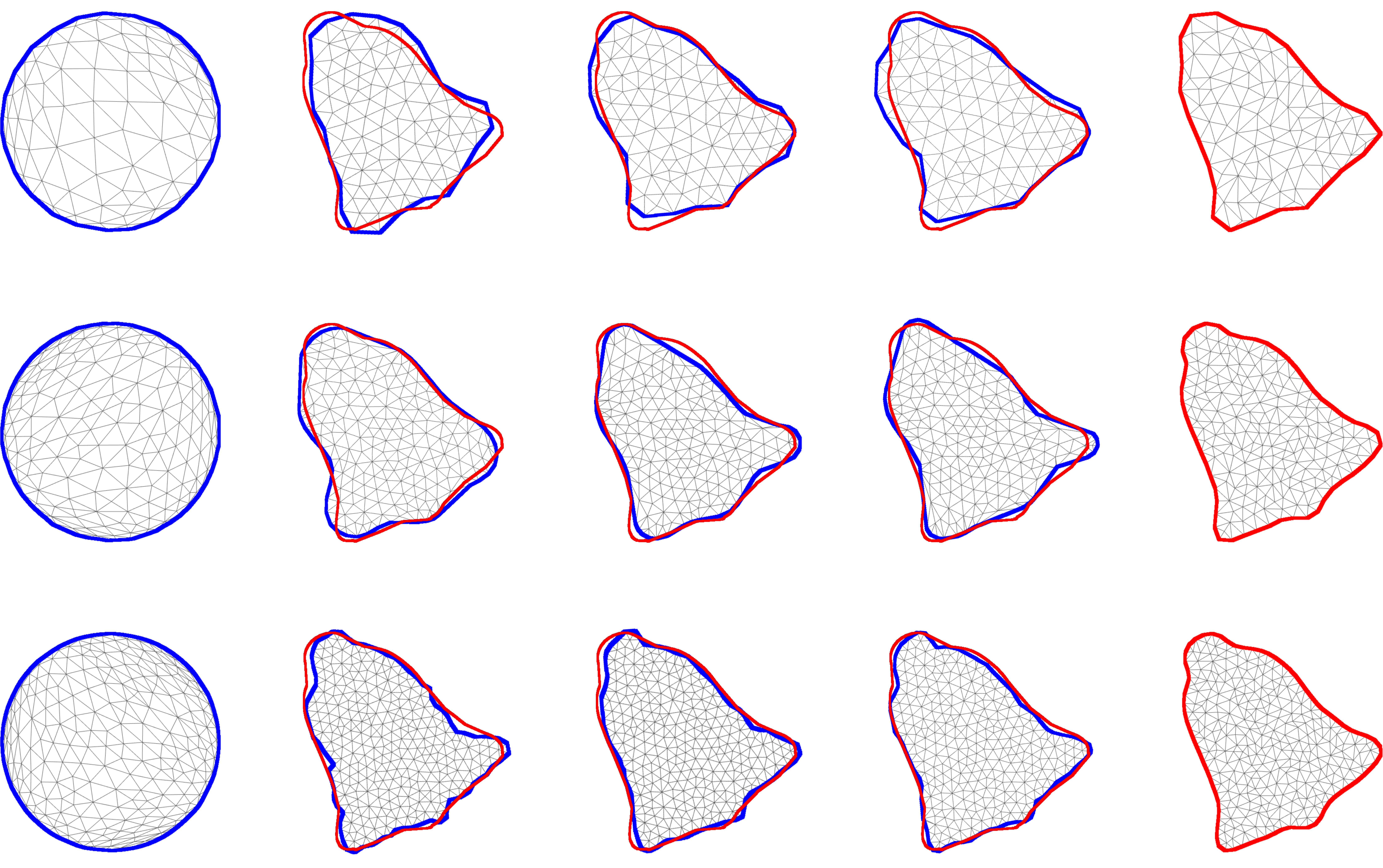}
%\put(-5,65){\footnotesize $k=10$}
\put(5,66){\footnotesize Init.}
\put(23,66){\footnotesize $k=10$}
\put(44,66){\footnotesize $k=20$}
\put(65,66){\footnotesize $k=30$}
\put(86,66){\footnotesize Target}
\put(29,45){\footnotesize $0.90$}
\put(49,45){\footnotesize $0.90$}
\put(70,45){\footnotesize $0.90$}
\put(29,23){\footnotesize $0.92$}
\put(49,23){\footnotesize $0.92$}
\put(70,23){\footnotesize $0.91$}
\put(29,0){\footnotesize $0.93$}
\put(49,0){\footnotesize $0.94$}
\put(70,0){\footnotesize $0.94$}
\end{overpic}
\vspace{0.1cm}
  \caption{\label{fig:bells}Shape recovery at increasing mesh resolution ($n=100$, $200$, and $300$ vertices increasing top to bottom) and bandwidth ($k=10$, $20$, and $30$ eigenvalues increasing left to right). In each test, the mesh graph {\em connectivity} of the target is known and input to the optimization process. We report the IOU score below each reconstruction. A finer sampling significantly improves reconstruction quality, whereas extending the bandwidth above $k=30$ does not lead to further improvement.}
\end{figure}
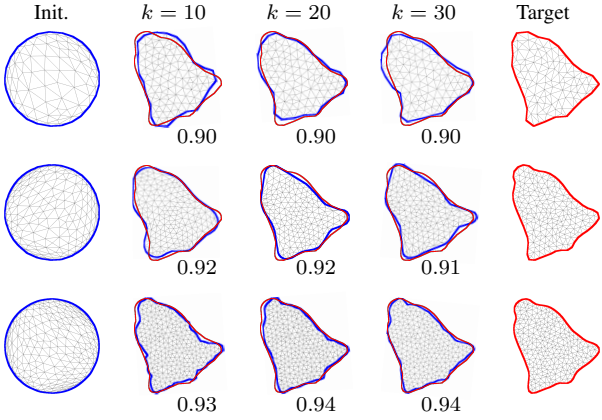

Using a standard $\ell_2$ norm for the data term in~\eqref{eq:p1} would not lead to accurate shape recovery: Since the high end of the spectrum accounts for small geometric variations of the embedding, a local optimum can be reached by perfectly aligning the high frequencies and concentrating most of the alignment error on the lower end (which accounts for the more global shape appearance). To make error diffusion more balanced, we thus adopt the weighted norm
\begin{equation}
\| \bm{\lambda} - \bm{\mu}  \|_\omega = \sum_{i=1}^k  \frac{1}{i} (\lambda_i-\mu_i)^2 \,.
\end{equation}
%
%which downweighs the high frequencies.

Problem~\eqref{eq:p1} seeks a Euclidean embedding whose Laplacian eigenvalues align to the ones given as input. This problem is highly non-linear and thus particularly difficult, %since 1) the optimization variables are nested, and 2) the problem is highly non-linear and thus
making it susceptible to local minima. 
Nevertheless, in all our tests we observed almost perfect alignment; we refer to the implementation Section~\ref{sec:opt} for further details.

%In this paper we consider flat shapes ($d=2$) as well as surfaces ($d=3$), as described in the following.

\subsection{Flat shapes}
When the embedding space is $\mathbb{R}^2$, a shape $\X$ is entirely determined by its boundary $\partial\X$. For this reason, we consider a variant of Problem~\eqref{eq:p1} where we optimize only for the boundary vertices.% and keep the interior vertices untouched.

\vspace{1ex}\noindent\textbf{Regularizers.}
We adopt the composite penalty
\begin{equation}
\rho_X(\mathbf{V}) = \rho_{X,1}(\mathbf{V}) + \rho_{X,2}(\mathbf{V})\,,
\end{equation}
where $\rho_{X,1}(\mathbf{V})$ is a Tikhonov regularizer promoting short edge lengths and thus a uniformly sized mesh:
\begin{equation}
\rho_{X,1}(\mathbf{V}) = \sum_{e_{ij}\in E_b} \ell^2_{ij}(\mathbf{V})\,,
\end{equation}
and $\rho_{X,2}(\mathbf{V})$ is defined as:
%
%\begin{equation}
%\rho_{X,2}(\mathbf{V}) = \sum_{ijk \in F} \bigl(\begin{smallmatrix}0\\0\\1\end{smallmatrix}\bigr)^\top ( (\mathbf{v}^j - \mathbf{v}^i)\times(\mathbf{v}^k - \mathbf{v}^i))
%\end{equation}
%
\begin{equation}
\rho_{X,2}(\mathbf{V}) = (\sum_{ijk \in F}{ (\mathbf{R}_{\tfrac{\pi}{2}}(\mathbf{v}^j - \mathbf{v}^i))^\top(\mathbf{v}^k - \mathbf{v}^i)})_{-}\,,
\end{equation}
where $\mathbf{R}_{\tfrac{\pi}{2}}=\bigl(\begin{smallmatrix}0&-1\\1&0\end{smallmatrix}\bigr)$ rotates 2D vectors by $\frac{\pi}{2}$ and $(x)_- = (\min\{0,x\})^2$. 
This term penalizes triangle flips that may occur throughout the optimization, and %, hence yielding an invalid embedding of the manifold.
works under the assumption of clockwise oriented triangles.

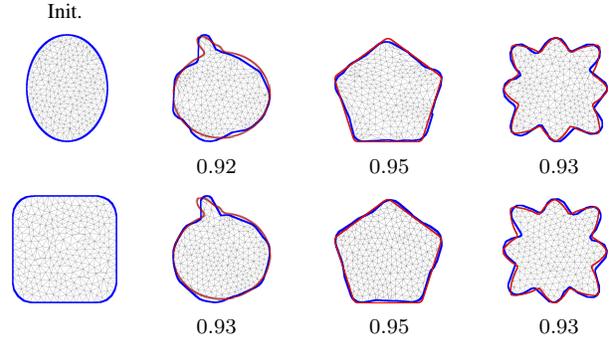
\begin{figure}[t!]
  \centering
\begin{overpic}
[trim=0cm 0cm 0cm 0cm,clip,width=0.95\linewidth]{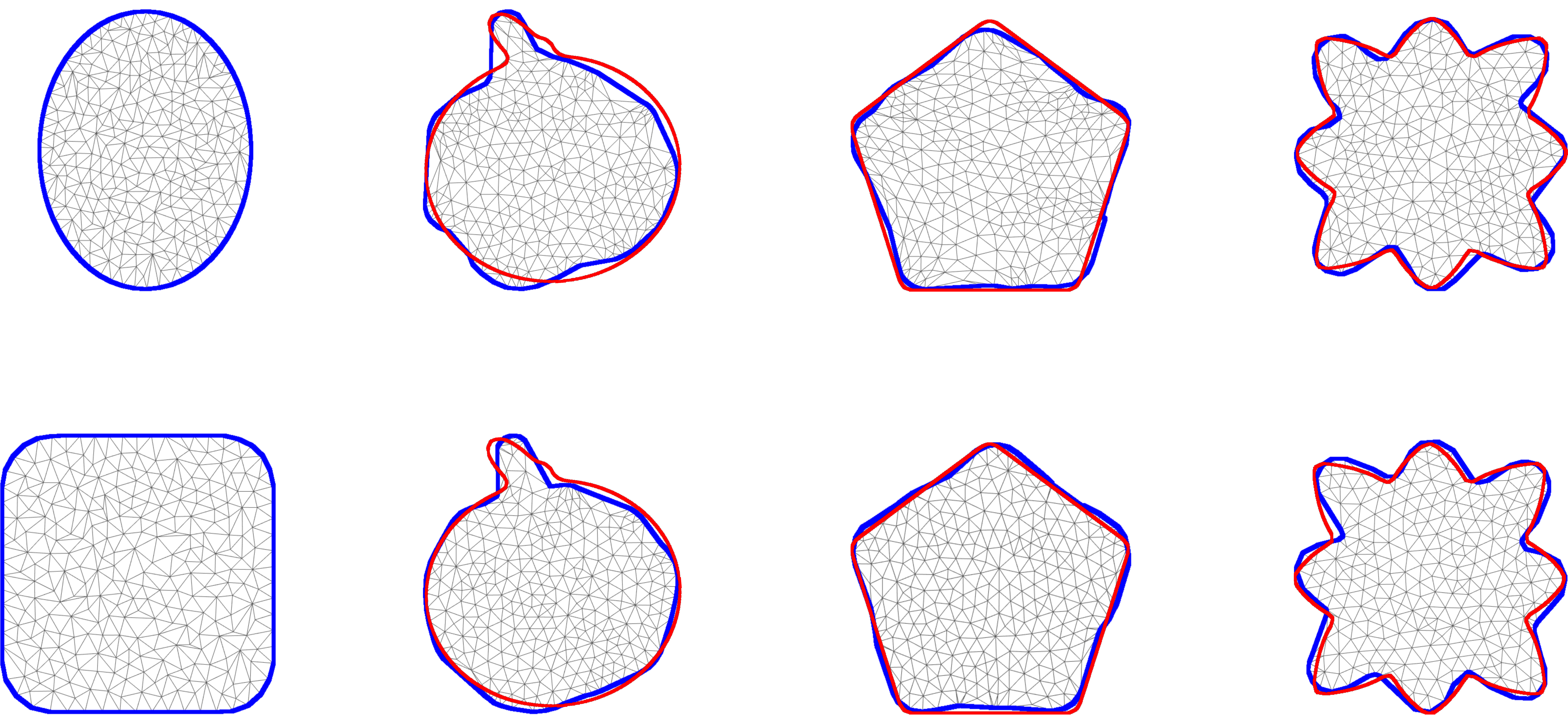}
\put(6,48){\footnotesize Init.}
\put(31,22){\footnotesize $0.92$}
\put(60,22){\footnotesize $0.95$}
\put(88.5,22){\footnotesize $0.93$}
\put(31,-5){\footnotesize $0.93$}
\put(60,-5){\footnotesize $0.95$}
\put(88.5,-5){\footnotesize $0.93$}
\end{overpic}
\vspace{0.5cm}
  \caption{\label{fig:init2d}Shape recovery with unknown mesh connectivity, under two different initializations. We report the IOU score below each recovered embedding. Differently from Figure~\ref{fig:bells}, in these tests the mesh tessellation was chosen arbitrarily and is in no way related to the input sequence of eigenvalues.}
\end{figure}

\vspace{2ex}\noindent\textbf{Error measure.}
We quantify the reconstruction quality as the area ratio of the intersection of the recovered and target embeddings over their union (IOU, the higher the better) after an optimal alignment has been carried out. In our plots, we visualize the recovered embedding with a blue outline, and the ground-truth (unknown) target with a red outline.

\vspace{2ex}\noindent\textbf{Mesh resolution and bandwidth.}
By operating with a discrete Laplace operator, our optimization problem is directly affected by the quality of the discretization. We investigate this dependency by running an evaluation at varying mesh resolution (in terms of number of vertices) and spectral bandwidth (number $k$ of input eigenvalues). The results are reported in Figure~\ref{fig:bells}. 

\vspace{2ex}\noindent\textbf{Examples.}
In Figure~\ref{fig:init2d} we show additional reconstruction results for different shapes. 
We remark that, differently from the previous tests, in these experiments we do {\em not} assume the mesh connectivity to be known. This way we put ourselves in the most general setting where the only input information is represented by the eigenvalues, thus factoring out any geometric aid that might be implicitly encoded in the connectivity graph.

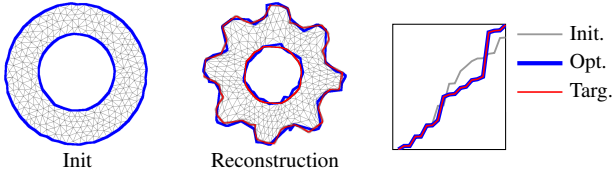
\begin{figure}[t]
  \centering
%\vspace{0.3cm}
\begin{overpic}
[trim=0cm 0cm 0cm 0cm,clip,width=0.55\linewidth]{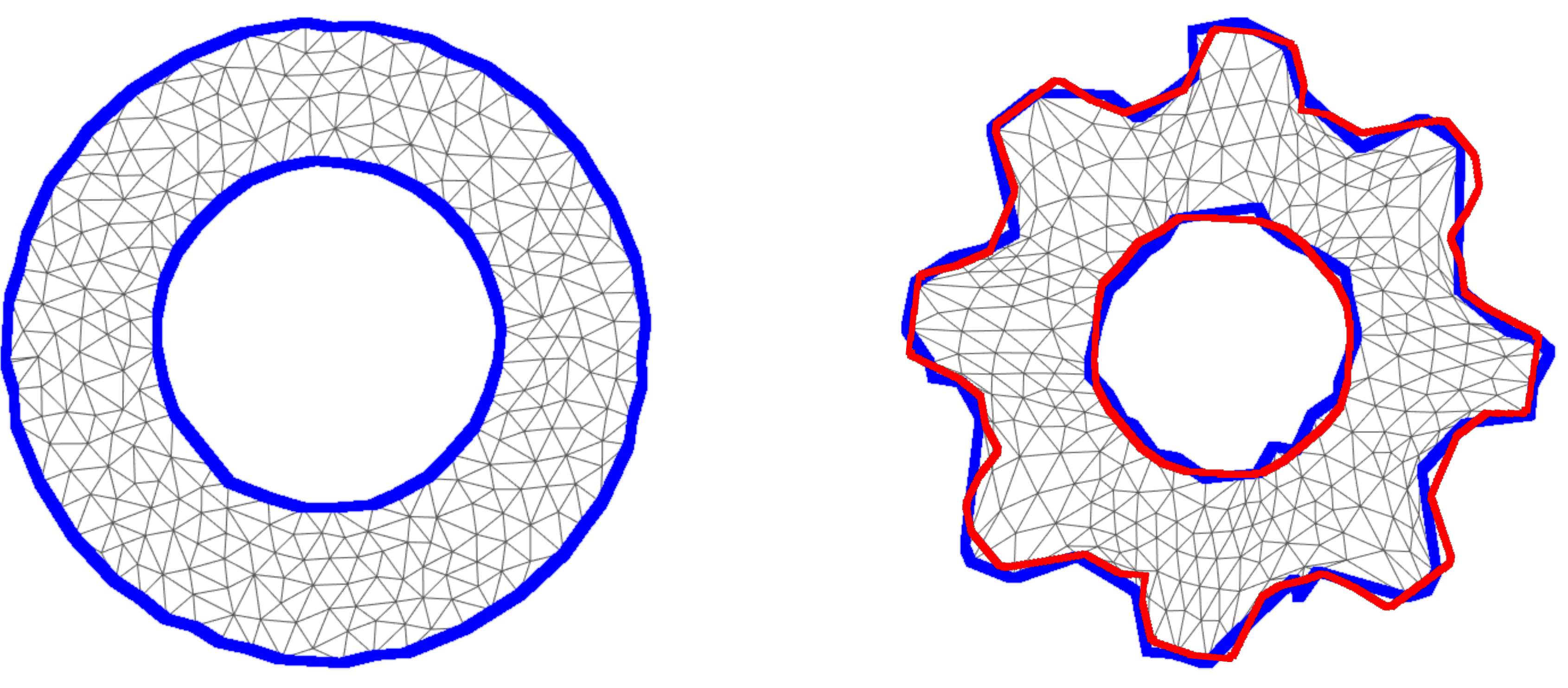}
\put(17,-5){\footnotesize Init}
\put(60,-5){\footnotesize Reconstruction}
%\put(58,23){\footnotesize Target}
%\put(83.5,23){\footnotesize Overlaid}
\end{overpic}
\hspace{0.4cm}
\definecolor{mycolor1}{rgb}{0.6,0.6,0.6}%
\definecolor{mycolor2}{rgb}{0,0,1}%
\definecolor{mycolor3}{rgb}{1,0,0}%
\begin{tikzpicture}

\begin{axis}[%
width=0.18\linewidth,
height=0.2\linewidth,
at={(0.758in,0.481in)},
scale only axis,
xmin=0,
xmax=20,
ymin=0,
ymax=160,
ticks=none,
legend style={legend cell align=left, align=left, draw=white!15!white},
legend style={at={(1.01,0.69)},anchor=west},
axis background/.style={fill=white}
]

\addplot [color=mycolor1,  line width=0.7pt]
  table[row sep=crcr]{%
1	2.4755e-14\\
2	3.7512\\
3	3.8243\\
4	14.823\\
5	14.969\\
6	32.522\\
7	32.757\\
8	55.824\\
9	56.327\\
10	84.003\\
11	84.502\\
12	100.05\\
13	106.32\\
14	112.82\\
15	116.27\\
16	116.51\\
17	122.24\\
18	123.3\\
19	142.44\\
20	143.07\\
};
\addlegendentry{\footnotesize Init.}

\addplot [color=mycolor2,  line width=1.6pt]
  table[row sep=crcr]{%
1	1.2461e-15\\
2	4.5435\\
3	4.552\\
4	16.537\\
5	16.677\\
6	30.465\\
7	30.826\\
8	37.44\\
9	67.271\\
10	70.275\\
11	70.971\\
12	78.628\\
13	80.2\\
14	88.549\\
15	89.777\\
16	92.983\\
17	149.58\\
18	151.75\\
19	152.2\\
20	159.09\\
};
\addlegendentry{\footnotesize Opt.}

\addplot [color=mycolor3, line width=0.6pt]
  table[row sep=crcr]{%
1	1.1348e-15\\
2	4.5614\\
3	4.6029\\
4	16.558\\
5	16.667\\
6	30.441\\
7	30.822\\
8	37.388\\
9	67.272\\
10	70.132\\
11	71.052\\
12	78.659\\
13	80.173\\
14	88.66\\
15	89.823\\
16	93.044\\
17	149.56\\
18	151.74\\
19	152.3\\
20	159.12\\
};
\addlegendentry{\footnotesize Targ.}

\end{axis}
\end{tikzpicture}%
\vspace{0.01cm}
  \caption{\label{fig:gear}Reconstruction of a non-simply connected shape. On the right we also show the initial, optimized, and target spectra.} %(30 evals 300 points).}
\end{figure}

\vspace{2ex}\noindent\textbf{Topology.}
At no point in our pipeline we assume the manifolds to be simply-connected. An example of recovery of a shape with a hole is given in Figure~\ref{fig:gear}. We do assume, however, to know the topological class (\eg, annulus-like rather than disc-like) of the target.

\begin{figure}[b!]
  \centering
\begin{overpic}
[trim=0cm 0cm 0cm 0cm,clip,width=0.9\linewidth]{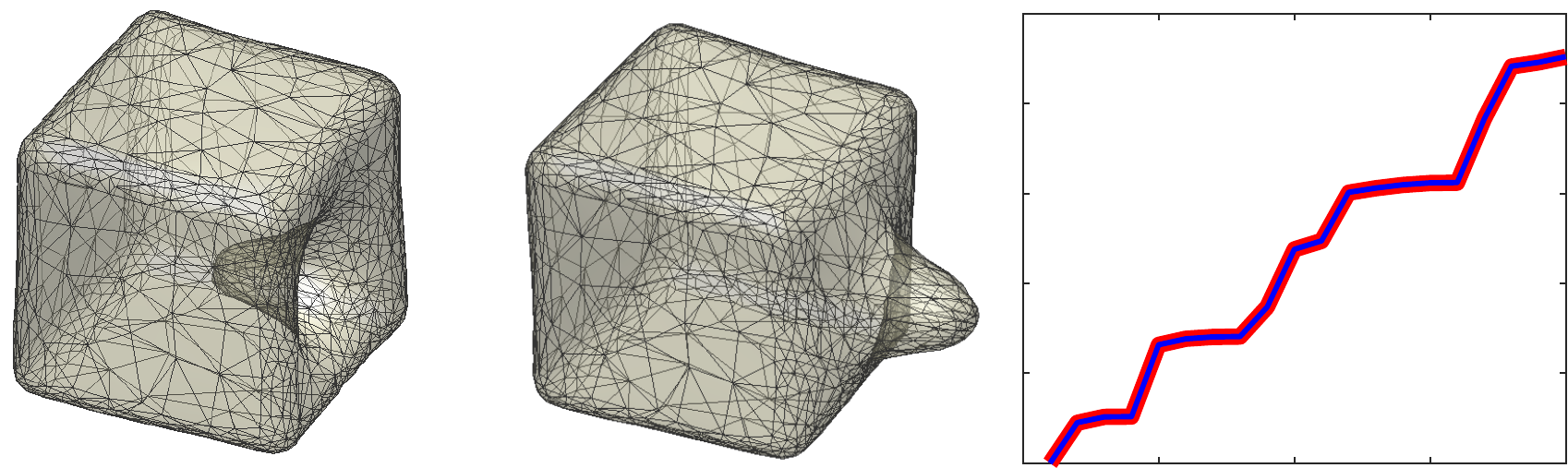}
\end{overpic}
  \caption{\label{fig:cube}Two isometric shapes with different volume; their (identical) spectra are shown on the right. The volume regularizer \eqref{eq:vol} allows to disambiguate the two solutions.}
\end{figure}

\subsection{Surfaces}
In the more general case of embeddings in $\mathbb{R}^3$ we once again adopt a composite penalty $\rho_X(\mathbf{V}) = \rho_{X,1}(\mathbf{V}) + \rho_{X,2}(\mathbf{V})$ with two different regularizers.

The first regularizer requires vertices to lie on the barycenter of their one-ring neighbors. It is defined as:
\begin{equation}
\rho_{X,1}(\mathbf{V}) = \|\mathbf{LV}\|_F^2\,,
\end{equation}
where $\mathbf{L}$ is the graph Laplacian of the initial embedding.
This term has the effect of promoting both a smooth surface and a more uniformly sampled embedding \cite{sorkine2004least}.

The second regularizer is a volume expansion term where the shape volume is estimated via the (discrete) divergence theorem as:
\begin{equation}\label{eq:vol}
\rho_{X,2}(\mathbf{V}) = - \bigl(\begin{smallmatrix}1\\1\\1\end{smallmatrix}\bigr)^\top \hspace{-0.1cm} \sum_{ijk \in F} \hspace{-0.1cm} ((\mathbf{v}^j-\mathbf{v}^i)\times(\mathbf{v}^k-\mathbf{v}^j))(\mathbf{v}^i+\mathbf{v}^j+\mathbf{v}^k)
\end{equation}
This term is useful in disambiguating isometries which differ by a change in volume (see Figure~\ref{fig:cube}).

\subsection{Implementation details}\label{sec:opt}

\begin{figure}[tb]
  \includegraphics[width=\linewidth]{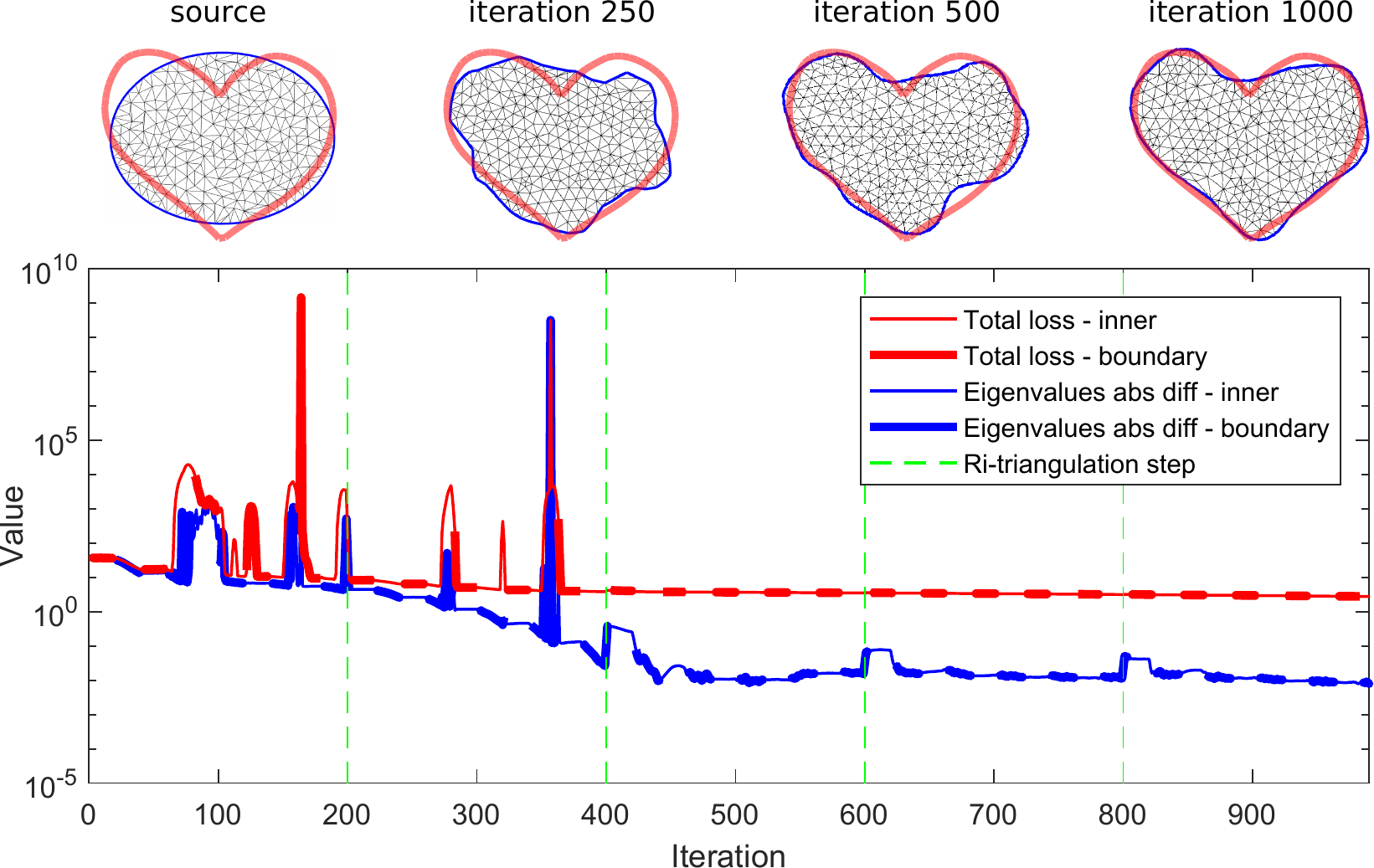}
  \caption{Energy plot during our alternating optimization process for flat shapes. Spikes in the energy are due triangle flips; the optimization procedure is able to recover from such cases and reaches a stable minimum that is close to a global optimum.}
  \label{fig:energy_2d}
\end{figure}

\begin{figure*}[t]
  \centering
\begin{overpic}
[trim=0cm -0.1cm 0cm 0cm,clip,width=1\linewidth]{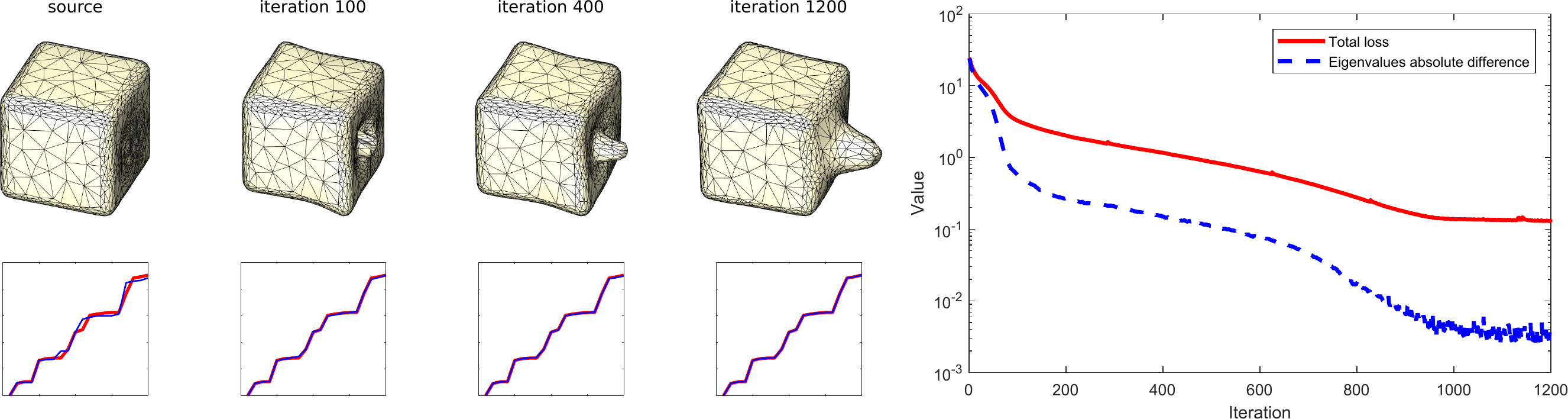}
\end{overpic}
  \caption{\label{fig:energy}An example of the reconstruction process; the target is a cube with a bump similar to the rightmost embedding. The plots under each shape show the current eigenvalues alignment (the target is the blue curve). Observe the staircase-like pattern due to the symmetry of the cube. On the right we show the evolution of the total energy (red curve) and the residual of eigenvalue alignment (blue curve).}
\end{figure*}
\begin{figure*}[t]
  \centering
%\vspace{0.cm}
\begin{overpic}
[trim=-1cm 0cm 0cm 0cm,clip,width=0.92\linewidth]{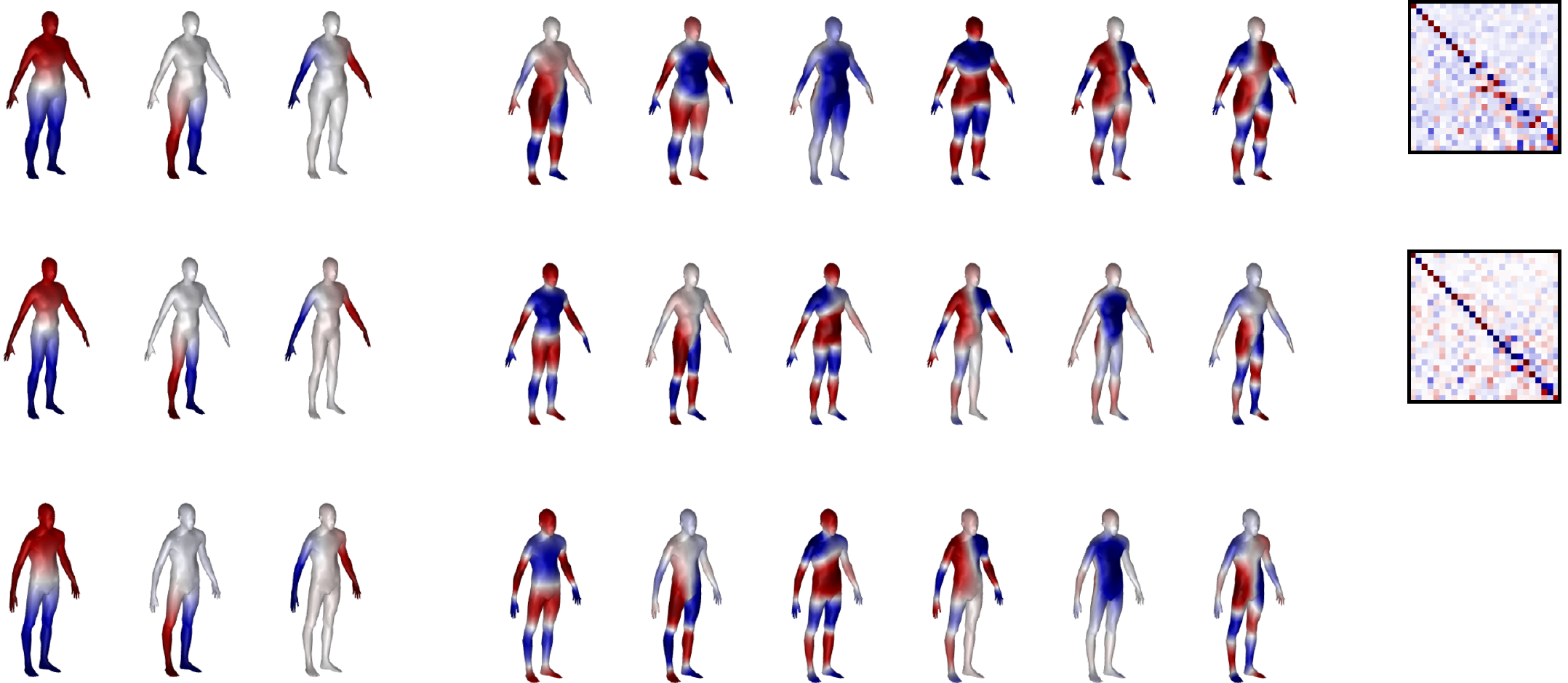}
\put(29,35){\Large{$\mathbb{\cdots}$}}
\put(29,20){\Large{$\mathbb{\cdots}$}}
\put(29,5){\Large{$\mathbb{\cdots}$}}
\put(1,35){\footnotesize {$\mathcal{X}$}}
\put(1,20){\footnotesize {$\mathcal{X'}$}}
\put(1,5){\footnotesize {$\mathcal{Y}$}}

\put(6,-2.5){\footnotesize $\phi_1$}
\put(15,-2.5){\footnotesize $\phi_2$}
\put(24,-2.5){\footnotesize $\phi_3$}
\put(37,-2.5){\footnotesize $\phi_{11}$}
\put(45.5,-2.5){\footnotesize $\phi_{12}$}
\put(54,-2.5){\footnotesize $\phi_{13}$}
\put(62.5,-2.5){\footnotesize $\phi_{14}$}
\put(71,-2.5){\footnotesize $\phi_{15}$}
\put(79.5,-2.5){\footnotesize $\phi_{16}$}

\put(6,13.5){\footnotesize $\psi_1'$}
\put(15,13.5){\footnotesize $\psi_2'$}
\put(24,13.5){\footnotesize $\psi_3'$}
\put(37,13.5){\footnotesize $\psi_{11}'$}
\put(45.5,13.5){\footnotesize $\psi_{12}'$}
\put(54,13.5){\footnotesize $\psi_{13}'$}
\put(62.5,13.5){\footnotesize $\psi_{14}'$}
\put(71,13.5){\footnotesize $\psi_{15}'$}
\put(79.5,13.5){\footnotesize $\psi_{16}'$}

\put(6,28.5){\footnotesize $\psi_1$}
\put(15,28.5){\footnotesize $\psi_2$}
\put(24,28.5){\footnotesize $\psi_3$}
\put(37,28.5){\footnotesize $\psi_{11}$}
\put(45.5,28.5){\footnotesize $\psi_{12}$}
\put(54,28.5){\footnotesize $\psi_{13}$}
\put(62.5,28.5){\footnotesize $\psi_{14}$}
\put(71,28.5){\footnotesize $\psi_{15}$}
\put(79.5,28.5){\footnotesize $\psi_{16}$}

\put(91.5,30){\footnotesize $\langle \psi_i, T \phi_i \rangle$}
\put(91.5,14.5){\footnotesize $\langle \psi_i', T \phi_i \rangle$}
\end{overpic}
\vspace{0.5cm}
  \caption{\label{fig:eigenfunctions}Alignment of eigenspaces as a result of isospectralization for non-isometric shape matching. Starting from a source shape $\X$ (first row), our algorithm solves for a new embedding $\X'$ (middle row) having the same Laplacian eigenvalues as those of the target $\Y$. Note how $\X'$ has the pose of $\X$, but the style of $\Y$. Remarkably, the eigenvalue alignment induces an alignment of the corresponding eigenfunctions, making the pairs $(\psi'_i,\phi_i)$ more similar than the initial pairs $(\psi_i,\phi_i)$. This is reflected in more diagonal functional map matrices (rightmost column), which in turn leads to a better conditioning for shape matching algorithms.}
\end{figure*}
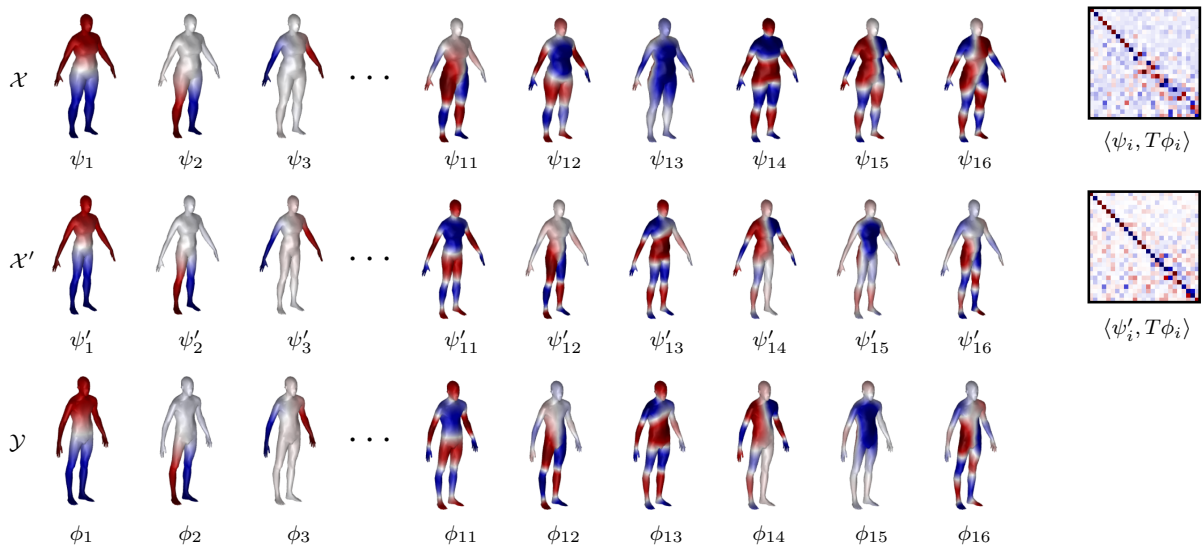
%

%In implementing the described optimization there are some technicalities that need to be taken into account, mostly due to ugly triangulations of the embedding that could occur during the minimization.

In the flat shape scenario, even though we optimize \eqref{eq:p1} only over boundary vertices, the interior vertices need to be moved as well so as to maintain a regular sampling. We do so by alternating optimization: First, the boundary vertices are updated for 10 iterations; then, interior vertices are re-positioned by minimizing their induced squared edge length. To avoid degenerate triangles, after 200 steps we also recompute a new triangulation from scratch while keeping the boundary edges fixed. While this procedure has no convergence guarantees, we observed convergence in all our experiments. An example of energy behavior during minimization is shown in Figures~\ref{fig:energy_2d} and \ref{fig:energy}.

For the numerical optimization we leverage auto-differentiation, and employ the Adam~\cite{adam} optimizer of Tensorflow~\cite{tensorflow2015}, adopting a cosine decay strategy for the regularizer weights. %For flat shapes, we alternate optimization over boundary and interior points every 10 iterations, while a resampling step is performed once every 200 iterations.
 Unless otherwise specified, we only use the first 30 eigenvalues on both flat shapes and surfaces, resampled respectively to 400 and 1000 points.

We finally note that our optimization strategy has no guarantee to reach a (local or global) optimum; however, in all our tests we empirically observed negligible numerical residual after eigenvalues alignment.

%\paragraph{Surface optimization}
%To attenuate the sensitivity of eigenvalues to noise and sampling we add a regularizer to the optimization problem. The first regularizer accounts for an as uniform as possible sampling of the shape, promoting a vertex to lie in the barycenter of its one ring neighbourhood. Let $A$ be the adjacency matrix of $N$, we can define the deviation from a uniform sampling penalty as:
%\begin{equation}
%\begin{array}{rcl}
%	B &=& A (I A\vec{1})^{-1} - I\\
%	\rho_{u} &=& \|B\delta_X^N\|_2^2
%\end{array}	 
%\end{equation}
%Note that this term is applied to the displacement field, having the effect of favour a smooth displacement of the initial points.   

% with $\alpha_1=1e6, \alpha_2=1e0, \alpha_3=1e0$.

%
\begin{figure*}[t]
  \centering
%\vspace{0.cm}
\definecolor{mycolor1}{rgb}{0.00000,0.44700,0.74100}%
\definecolor{mycolor2}{rgb}{0.85000,0.32500,0.09800}%
\begin{tikzpicture}

\begin{axis}[%
width=0.2\linewidth,
height=0.2\linewidth,
scale only axis,
xmin=0,
xmax=0.5,
xlabel style={font=\color{white!15!black}},
xlabel={\footnotesize Geodesic error},
ymin=0,
ymax=100,
ylabel style={font=\color{white!15!black}},
ylabel={\footnotesize \% Correspondences},
axis background/.style={fill=white},
axis x line*=bottom,
axis y line*=left,
xmajorgrids,
ymajorgrids,
every x tick label/.append style={font=\color{black}, font=\tiny},
every y tick label/.append style={font=\color{black}, font=\tiny},
xtick={0,0.1,0.2,0.3,0.4,0.5},
xticklabels={0,0.1,0.2,0.3,0.4,0.5},
ylabel style={at={(0.155,0.48)}},
xlabel style={at={(0.5,0.07)}},
legend style={legend cell align=left, align=left, draw=white!15!black, at={(0.44,0.1)}, anchor= south west}
]
\addplot [color=mycolor1, line width=2.0pt]
  table[row sep=crcr]{%
0	1.9873\\
0.0100000000000051	2.0397\\
0.019999999999996	3.1889\\
0.0300000000000011	5.7\\
0.0400000000000063	8.8349\\
0.0499999999999972	12.635\\
0.0600000000000023	17.168\\
0.0699999999999932	22.135\\
0.0799999999999983	27.697\\
0.0900000000000034	33.763\\
0.0999999999999943	39.986\\
0.109999999999999	45.454\\
0.120000000000005	51.273\\
0.129999999999995	55.967\\
0.140000000000001	60.175\\
0.150000000000006	64.013\\
0.159999999999997	67.262\\
0.170000000000002	70.313\\
0.180000000000007	72.77\\
0.189999999999998	75.165\\
0.200000000000003	77.135\\
0.209999999999994	79.073\\
0.219999999999999	80.868\\
0.230000000000004	82.162\\
0.239999999999995	83.427\\
0.25	84.595\\
0.260000000000005	85.519\\
0.269999999999996	86.403\\
0.280000000000001	87.198\\
0.290000000000006	87.91\\
0.299999999999997	88.525\\
0.310000000000002	89.084\\
0.329999999999998	89.998\\
0.340000000000003	90.4\\
0.349999999999994	90.722\\
0.359999999999999	91.022\\
0.370000000000005	91.41\\
0.379999999999995	91.754\\
0.390000000000001	92.092\\
0.400000000000006	92.44\\
0.409999999999997	92.694\\
0.420000000000002	92.897\\
0.439999999999998	93.408\\
0.450000000000003	93.6\\
0.459999999999994	93.835\\
0.469999999999999	94.089\\
0.480000000000004	94.273\\
0.489999999999995	94.373\\
0.5	94.462\\
};
\addlegendentry{\footnotesize Before}

\addplot [color=mycolor2, line width=2.0pt]
  table[row sep=crcr]{%
0	6.654\\
0.0100000000000051	6.75239999999999\\
0.019999999999996	9.6413\\
0.0300000000000011	16.2\\
0.0400000000000063	23.687\\
0.0499999999999972	31.546\\
0.0699999999999932	48.038\\
0.0799999999999983	55.902\\
0.0900000000000034	63.351\\
0.0999999999999943	69.606\\
0.109999999999999	74.708\\
0.120000000000005	79.148\\
0.129999999999995	82.595\\
0.140000000000001	85.421\\
0.150000000000006	87.74\\
0.159999999999997	89.483\\
0.170000000000002	90.925\\
0.180000000000007	92.049\\
0.189999999999998	93.027\\
0.200000000000003	93.922\\
0.209999999999994	94.637\\
0.219999999999999	95.254\\
0.230000000000004	95.789\\
0.239999999999995	96.237\\
0.25	96.617\\
0.260000000000005	96.954\\
0.269999999999996	97.254\\
0.280000000000001	97.524\\
0.290000000000006	97.76\\
0.299999999999997	97.987\\
0.310000000000002	98.205\\
0.319999999999993	98.387\\
0.329999999999998	98.576\\
0.340000000000003	98.703\\
0.349999999999994	98.808\\
0.359999999999999	98.906\\
0.370000000000005	98.987\\
0.379999999999995	99.052\\
0.390000000000001	99.102\\
0.400000000000006	99.144\\
0.409999999999997	99.163\\
0.420000000000002	99.176\\
0.430000000000007	99.181\\
0.439999999999998	99.184\\
0.459999999999994	99.194\\
0.480000000000004	99.216\\
0.489999999999995	99.221\\
0.5	99.237\\
};
\addlegendentry{\footnotesize After}

\end{axis}
\end{tikzpicture}%
\hspace{0.6cm}
\begin{overpic}
[trim=0cm -3cm 0cm 0cm,clip,width=0.65\linewidth]{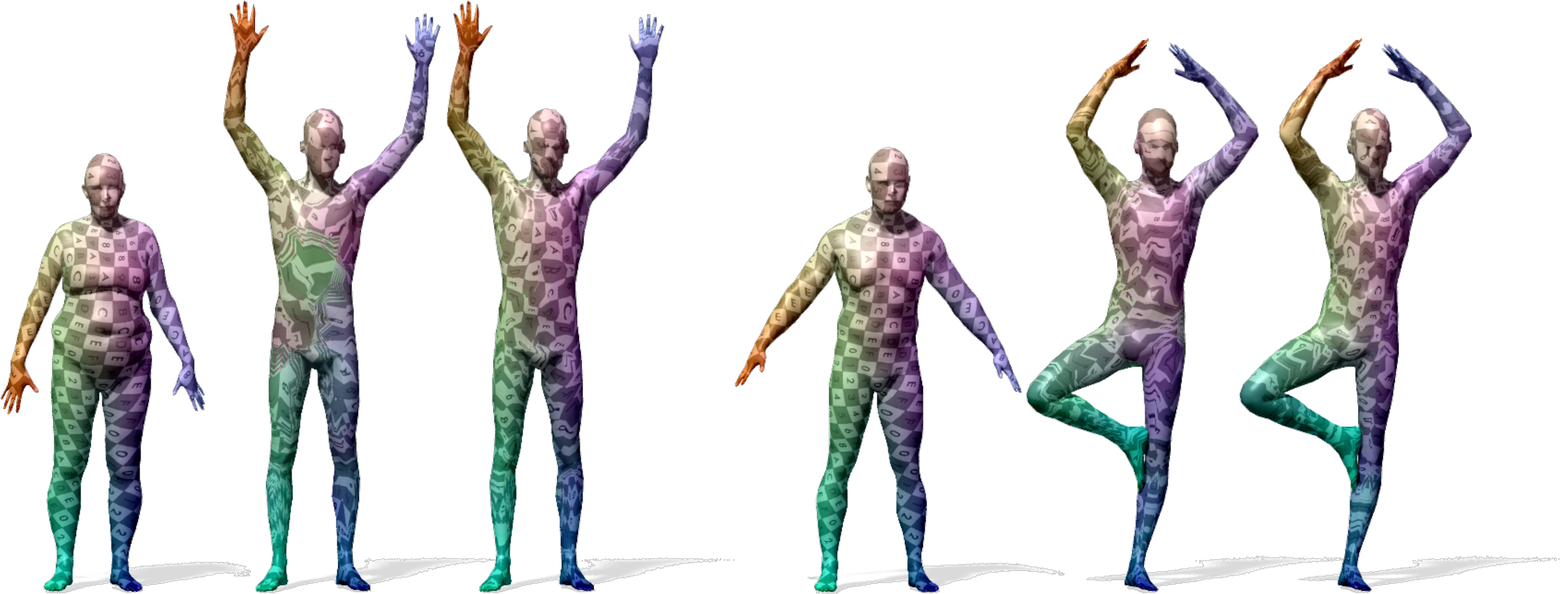}
\put(1,1){\footnotesize Reference}
\put(51,1){\footnotesize Reference}
\put(17,1){\footnotesize Before}
\put(17,-2){\footnotesize isospec.}
\put(32,1){\footnotesize After}
\put(32,-2){\footnotesize isospec.}
\put(68,1){\footnotesize Before}
\put(68,-2){\footnotesize isospec.}
\put(82,1){\footnotesize After}
\put(82,-2){\footnotesize isospec.}
\end{overpic}
\vspace{0.5cm}
  \caption{\label{fig:faust}Results on non-isometric shape matching. The plots on the left are averaged over 60 non-isometric shape pairs from the FAUST inter-subject dataset~\cite{faust}. In order to visualize correspondence, we use it to transfer a texture from reference shape to target.}
\end{figure*}
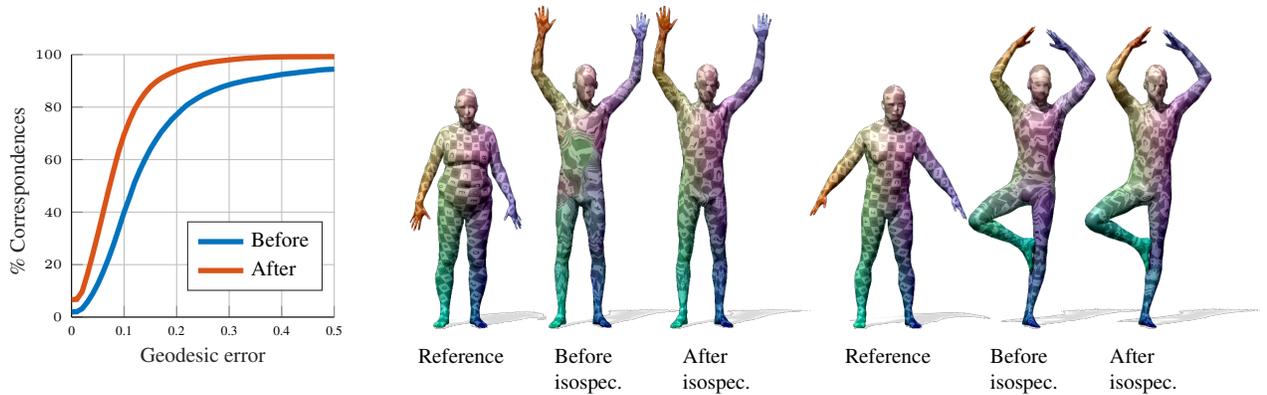

\section{Applications in shape analysis}

\subsection{Non-isometric shape matching}\label{sec:noniso}
We have applied our shape optimization approach to the problem of finding correspondences between non-rigid shapes. In this setting, we are given a pair of 3D shapes $\X, \Y$, both represented as triangle meshes, and our goal is to find a dense map $T: \X \rightarrow \Y$ between them. This is a very well-studied problem in computer vision and computer graphics, with a wide range of techniques proposed over the years (see \cite{van2011survey,tam2013registration,biasotti2016recent} for several surveys). 

Non-rigid shape matching is particularly difficult, as it would require designing a universal correspondence algorithm, capable of handling arbitrary deformations in a fully automatic way. A very successful sub-class of non-rigid shape deformations is \emph{intrinsic isometries}, in which the underlying map $T$ is assumed to approximately preserve \emph{geodesic distances} between pairs of points on the shapes. A large number of efficient methods has been proposed to solve the shape matching problem under this assumption~\cite{aflalo2016spectral,biasotti2016recent}. At the same time, most of these techniques result in very poor correspondences whenever the assumption of intrinsic isometry is not satisfied.

\vspace{1ex}\noindent\textbf{Approach.}
Our main insight is that the alignment of the spectra of two shapes can make them more intrinsically isometric, and thus can facilitate finding accurate correspondences using existing techniques. Given shapes $\X, \Y$, with Laplacian spectra $\bm{\lambda}_\X, \bm{\lambda}_\Y$ we propose to find a map between them using the following three-step approach:
\begin{enumerate}
  \setlength\itemsep{0em}
\item Deform $\X$  to obtain $\X'$ whose spectrum $\bm{\lambda}_{\X'}$ is better aligned with $\bm{\lambda}_\Y$.
\item Compute the correspondence $T' : \X' \rightarrow \Y$ using an existing isometric shape matching algorithm.
\item Convert $T'$ to $T: \X \rightarrow \Y$ using the identity map between $\X$ and $\X'$.
\end{enumerate}
Our main intuition is that as mentioned above, despite the existence of exceptional counter-examples, in most practical cases this procedure is very likely to make shapes $\X'$ and $\Y$ close to being isometric. Therefore, we would expect an isometric shape matching algorithm to match $\Y$ to $\X'$ better than to the original shape $\X$. Finally, after computing a map $T: \X' \rightarrow \Y$, we can trivially convert it to a map, since $\X,\X'$ are in 1-1 correspondence.

The approach described above builds upon the remarkable observation that aligning the Laplacian eigenvalues also induces an alignment of the {\em eigenspaces} of the two shapes. This is illustrated on a real example in Figure~\ref{fig:eigenfunctions}, where we show a subset of eigenfunctions for two non-isometric surfaces (a man and a woman) before and after isospectralization. In a sense, isospectralization implements a notion of {\em correspondence-free} alignment of the functional spaces spanned by the first $k$ Laplacian eigenfunctions.

\vspace{1ex}\noindent\textbf{Implementation.}
For this and the following application, we replaced the optimization variables by optimizing over a {\em displacement field} rather than the absolute vertex positions in Problem~\eqref{eq:p1}. Doing so, we observed a better quality of the recovered embeddings.

We have implemented the approach described above by using an existing shape correspondence algorithm \cite{nogneng2017informative} based on the functional maps framework \cite{ovsjanikov2012functional,ovsjanikov2017computing}. One of the advantages of this approach is that it is {\em purely intrinsic} and only depends on the quantities derived from the Laplace-Beltrami operators of the two shapes. 

\vspace{1ex}\noindent\textbf{Remark.} The exact embedding of the optimized shape $\X'$ does not play a role, and can be different from that of $\Y$. In other words, we do not aim to reproduce the shape $\Y$, but rather only use our shape optimization strategy as an auxiliary step to facilitate shape correspondence.

\begin{figure}[t]
  \centering
\begin{overpic}
[trim=0cm 0cm 0cm 0cm,clip,width=0.49\linewidth]{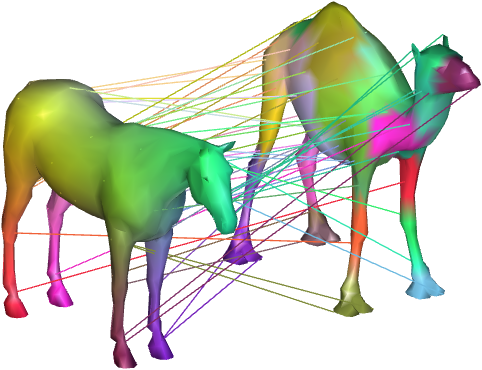}
\end{overpic}
\begin{overpic}
[trim=0cm 0cm 0cm 0cm,clip,width=0.49\linewidth]{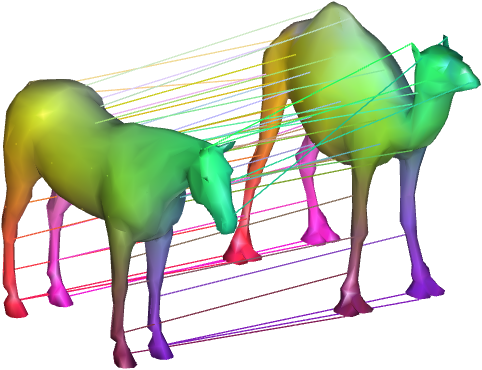}
\end{overpic}
  \caption{\label{fig:horse}Non-isometric shape matching before (left) and after (right) isospectralization. The correspondence is computed according to the algorithm of Section~\ref{sec:noniso}.}
\end{figure}

\begin{figure}[tb]
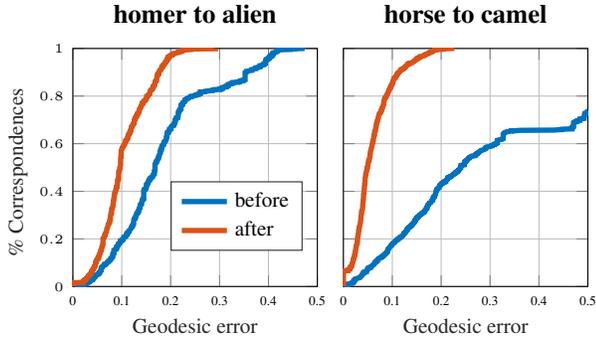

  \centering
\input{./homeralien_error.tikz}\hspace{-0.21cm}
\input{./horsecamel_error.tikz}
  \caption{\label{fig:nonisoq}Quantitative evaluation of non-isometric shape matching for the homer/alien pair of Figure~\ref{fig:teaser} and the horse/camel pair of Figure~\ref{fig:horse}. Isospectralization leads to a dramatic improvement in correspondence accuracy.}
\end{figure}

We use the functional maps-based algorithm of \cite{nogneng2017informative} with the open source implementation provided by the authors. This algorithm is based on first solving for a functional map represented in the Laplacian eigenbasis~\cite{aflalo2015optimality} by using several descriptor-preservation and regularization constraints, and then converting the functional map to a pointwise one. As done in \cite{nogneng2017informative}, we used the wave kernel signature \cite{wks} as descriptors and commutativity with the Laplace-Beltrami operators for map regularization. This leads to a convex optimization problem which can be efficiently solved with an iterative quasi-Newton method. Finally, we convert the functional map to a pointwise one using nearest neighbor search in the spectral domain as in \cite{ovsjanikov2012functional}. We evaluate the quality of the final correspondence by measuring the average geodesic error with respect to some externally-provided ground truth map \cite{kim2011blended}. We refer to Figures~\ref{fig:teaser} (bottom row), \ref{fig:faust}, \ref{fig:horse}, and \ref{fig:nonisoq} for quantitative and qualitative results.


\subsection{Style transfer}\label{sec:style}
\begin{figure}[t]
  \centering
%\vspace{0.6cm}
\begin{overpic}
[trim=0cm 0cm 0cm 0cm,clip,width=0.95\linewidth]{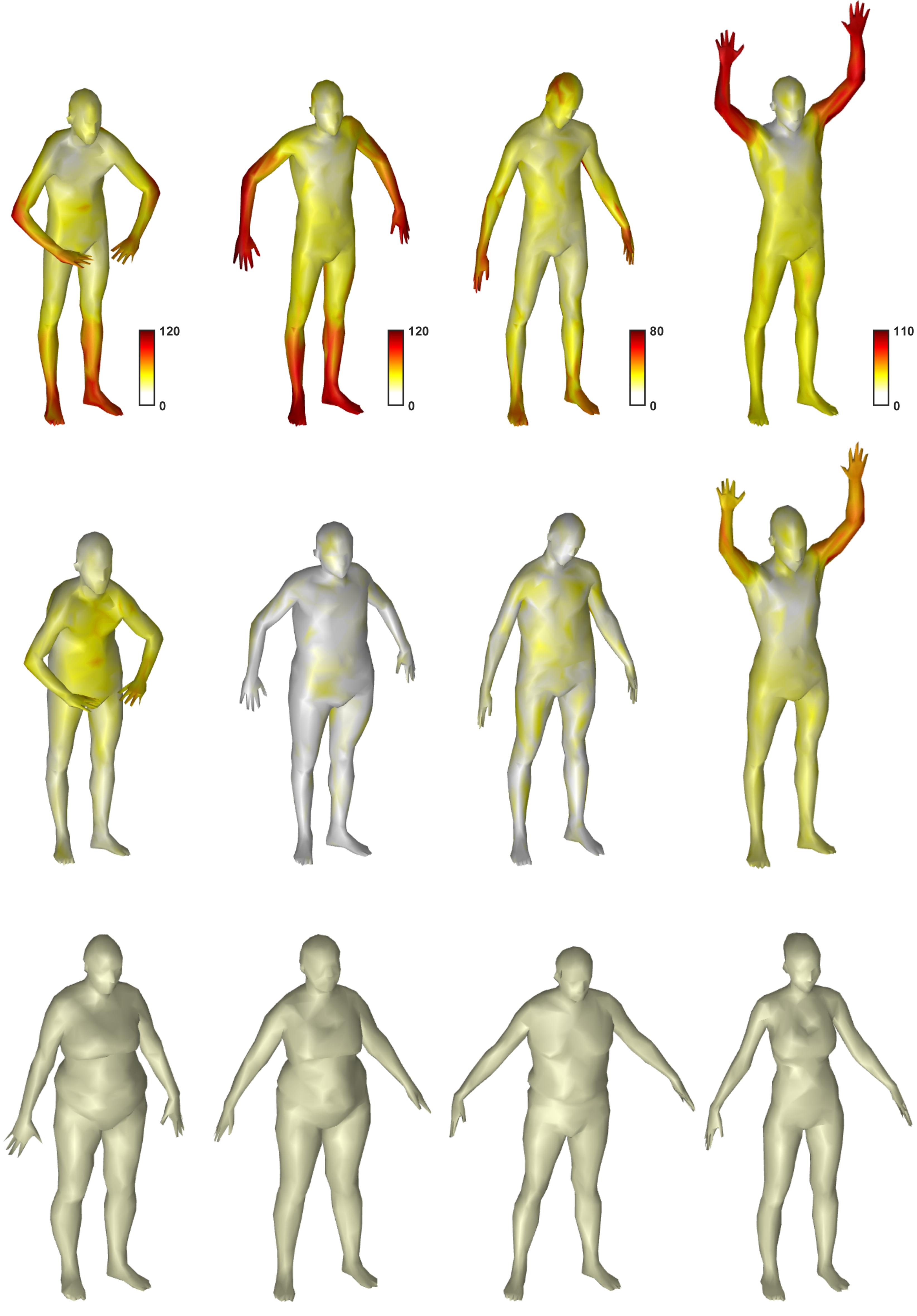}
\put(26,65){\footnotesize Source pose $\mathcal{X}$}
\put(19,30.75){\footnotesize Target style + source pose ($\X'$) }
\put(28,-1.5){\footnotesize Target style $\mathcal{Y}$}
\end{overpic}
\vspace{0.45cm}
  \caption{\label{fig:styletrans}Spectra alignment can be used for style transfer. In these four examples (one per column), we transfer the style of the target shape (third row) to the source shape (first row), obtaining the embeddings shown in the middle. Since here we use few eigenvalues, only smooth details are transferred. The colored heatmap encodes distortion of the geodesic distances $e(x_i)=\sum_{x_j\in A} \| d_A(x_i,x_j)-d_\mathcal{Y}(T(x_i),T(x_j))\|_2$, where $T$ is the ground truth map from $A=\{\X,\X'\}$ to $\mathcal{Y}$; it is higher on the limbs due to accumulation over larger distances.}
\end{figure}

As a second possible application we explore the task of style transfer between deformable shapes. Given a pair of surfaces $\X$ (the source) and $\Y$ (the target), the idea is to modify the geometric details of $\X$ to match those of shape $\Y$. %, while maintaining its overall pose in space as encoded in the lower-end portion of $\bm{\lambda}_\X$.
 We do so simply by recovering an embedding in $\mathbb{R}^3$ from the eigenvalues $\bm{\lambda}_\Y$, where we initialize the optimization with the source shape $\X$. Qualitative examples of this procedure are shown in Figure~\ref{fig:styletrans}.

\vspace{0.5ex}\noindent\textbf{Remark.} We emphasize that this way of transferring style among given shapes is completely {\em correspondence-free}, as it does {\em not} require a map between them. This is different from existing approaches like~\cite{levy06,boscaini2015shape}, which in addition to requiring the entire Laplacian matrix, also require a precise map between $\X$ and $\Y$ to be given as input.

\vspace{-0.1cm}

\section{Discussion and conclusions}
\vspace{-5mm}
\epigram{
\em ``Vibrations are everywhere, and so too are the eigenvalues associated with them.'' (Parlett, 1998)
}

In this paper, we addressed the decades-old problem of recovering a metric embedding from a (partial) measurement of its Laplacian eigenvalues by introducing a numerical procedure called isospectralization, trying to deform the shape embedding to align its Laplacian eigenvalues with a given spectrum. 
We find it remarkable that the use of isospectralization simplifies the problem of finding intrinsic correspondences between shapes, allowing us to significantly improve standard pipelines for shape matching almost for free. Interestingly, there is no \emph{a priori} guarantee that the isospectralization process deforms the shapes in a sensible way (for instance, it is conceivable that isospectralizing a horse into a camel would result in the horse's legs collapsing while a new set grows out of its back, which would spoil the construction of the correspondence).  
%In fact, one can construct optimization algorithms that could exhibit this feature. 
%As a simple example, consider an algorithm that tries to randomly deform the horse. It is conceivable that this approach could produce the undesirable behavior described above.
Our results suggest that such situations do not arise in practice when using a gradient descent-like algorithm that deforms the shape progressively.  
In other words, our approach to isospectralization deforms the shape in a meaningful way. It would be of interest to obtain a mathematically precise statement of this meaningfulness.
This would be of both practical importance in shape matching and, potentially, mathematical interest in the field of spectral geometry.

\vspace{-0.1cm}
{\small 
\section*{Acknowledgments}\vspace{-0.2cm}
\noindent The authors wish to thank Alex Bronstein for useful discussions.  ER and AR are supported by the ERC Starting Grant No. 802554 (SPECGEO). MB and LC are partially supported by ERC Consolidator Grant No. 724228 (LEMAN) and Google Research Faculty awards. MB is also partially supported by the Royal Society Wolfson Research Merit award and Rudolf Diesel industrial fellowship at TU Munich. Parts of this work were supported by a Google Focused Research Award, KAUST OSR Award No. OSR-CRG2017-3426, a gift from the NVIDIA Corporation and the ERC Starting Grant No. 758800 (EXPROTEA).
}

{\small
\bibliographystyle{ieee_fullname}
\bibliography{egbib}

\begin{thebibliography}{10}\itemsep=-1pt

\bibitem{aasen2013shape}
David Aasen, Tejal Bhamre, and Achim Kempf.
\newblock Shape from sound: toward new tools for quantum gravity.
\newblock {\em Physical Review Letters}, 110(12):121301, 2013.

\bibitem{tensorflow2015}
Mart\'{\i}n Abadi, Ashish Agarwal, Paul Barham, et~al.
\newblock {TensorFlow}: Large-scale machine learning on heterogeneous systems,
  2015.
\newblock Software available from tensorflow.org.

\bibitem{aflalo2015optimality}
Yonathan Aflalo, Haim Brezis, and Ron Kimmel.
\newblock On the optimality of shape and data representation in the spectral
  domain.
\newblock {\em SIAM Journal on Imaging Sciences}, 8(2):1141--1160, 2015.

\bibitem{aflalo2016spectral}
Yonathan Aflalo, Anastasia Dubrovina, and Ron Kimmel.
\newblock Spectral generalized multi-dimensional scaling.
\newblock {\em International Journal of Computer Vision}, 118(3):380--392,
  2016.

\bibitem{wks}
Mathieu Aubry, Ulrich Schlickewei, and Daniel Cremers.
\newblock The {W}ave {K}ernel {S}ignature: {A} {Q}uantum {M}echanical
  {A}pproach to {S}hape {A}nalysis.
\newblock In {\em Proc. ICCV Workshops}, 2011.

\bibitem{berger2012panoramic}
Marcel Berger.
\newblock {\em A panoramic view of Riemannian geometry}.
\newblock Springer Science \& Business Media, 2012.

\bibitem{bharaj2015computational}
Gaurav Bharaj, David~IW Levin, James Tompkin, Yun Fei, Hanspeter Pfister,
  Wojciech Matusik, and Changxi Zheng.
\newblock Computational design of metallophone contact sounds.
\newblock {\em TOG}, 34(6), 2015.

\bibitem{biasotti2016recent}
Silvia Biasotti, Andrea Cerri, Alex Bronstein, and Michael Bronstein.
\newblock Recent trends, applications, and perspectives in 3d shape similarity
  assessment.
\newblock {\em Computer Graphics Forum}, 35(6):87--119, 2016.

\bibitem{faust}
Federica Bogo, Javier Romero, Matthew Loper, and Michael~J Black.
\newblock {FAUST}: {D}ataset and {E}valuation for 3d {M}esh {R}egistration.
\newblock In {\em Proc. CVPR}, 2014.

\bibitem{borrelli2012flat}
Vincent Borrelli, Sa{\"\i}d Jabrane, Francis Lazarus, and Boris Thibert.
\newblock Flat tori in three-dimensional space and convex integration.
\newblock {\em PNAS}, 2012.

\bibitem{boscaini2015shape}
Davide Boscaini, Davide Eynard, Drosos Kourounis, and Michael~M Bronstein.
\newblock Shape-from-operator: Recovering shapes from intrinsic operators.
\newblock {\em Computer Graphics Forum}, 34(2):265--274, 2015.

\bibitem{chern18}
Albert Chern, Felix Kn\"{o}ppel, Ulrich Pinkall, and Peter Schr\"{o}der.
\newblock Shape from metric.
\newblock {\em TOG}, 37(4):63:1--63:17, 2018.

\bibitem{chu2005inverse}
Moody Chu and Gene Golub.
\newblock {\em Inverse eigenvalue problems: theory, algorithms, and
  applications}.
\newblock Oxford University Press, 2005.

\bibitem{corman2017functional}
Etienne Corman, Justin Solomon, Mirela Ben-Chen, Leonidas Guibas, and Maks
  Ovsjanikov.
\newblock Functional characterization of intrinsic and extrinsic geometry.
\newblock {\em TOG}, 36(2):14, 2017.

\bibitem{gordon1992isospectral}
Carolyn Gordon, David Webb, and Scott Wolpert.
\newblock Isospectral plane domains and surfaces via {R}iemannian orbifolds.
\newblock {\em Inventiones Mathematicae}, 110(1):1--22, 1992.

\bibitem{cannothear}
Carolyn Gordon, David~L. Webb, and Scott Wolpert.
\newblock One cannot hear the shape of a drum.
\newblock {\em Bulletin of the American Mathematical Society}, 27:134--138,
  1992.

\bibitem{hamidian16}
Hajar Hamidian, Jiaxi Hu, Zichun Zhong, and Jing Hua.
\newblock Quantifying shape deformations by variation of geometric spectrum.
\newblock In {\em Proc. MICCAI}, 2016.

\bibitem{hezari2010inverse}
Hamid Hezari and Steve Zelditch.
\newblock Inverse {S}pectral {P}roblem for {A}nalytic $(\mathbb{Z}/2
  \mathbb{Z})$-{S}ymmetric {D}omains in $\mathbb{R}^{N}$.
\newblock {\em Geometric and Functional Analysis}, 20(1):160--191, 2010.

\bibitem{hu17}
Jiaxi Hu, Hajar Hamidian, Zichun Zhong, and Jing Hua.
\newblock Visualizing shape deformations with variation of geometric spectrum.
\newblock {\em IEEE TVCG}, 23(1):721--730, 2017.

\bibitem{jacobson2012cotangent}
Alec Jacobson and Olga Sorkine-Hornung.
\newblock A cotangent laplacian for images as surfaces.
\newblock {\em Technical report/Department of Computer Science, ETH, Zurich},
  757, 2012.

\bibitem{kac.drum}
Mark Kac.
\newblock Can one hear the shape of a drum?
\newblock {\em The American Mathematical Monthly}, 73(4):1--23, 1966.

\bibitem{kempf2010spacetime}
Achim Kempf.
\newblock Spacetime could be simultaneously continuous and discrete, in the
  same way that information can be.
\newblock {\em New Journal of Physics}, 12(11):115001, 2010.

\bibitem{kim2011blended}
Vladimir~G Kim, Yaron Lipman, and Thomas Funkhouser.
\newblock {B}lended {I}ntrinsic {M}aps.
\newblock {\em TOG}, 30(4):79, 2011.

\bibitem{adam}
Diederik~P. Kingma and Jimmy Ba.
\newblock Adam: A method for stochastic optimization.
\newblock {\em CoRR}, abs/1412.6980, 2014.

\bibitem{levy06}
Bruno Levy.
\newblock Laplace-beltrami eigenfunctions towards an algorithm that
  "understands" geometry.
\newblock In {\em Proc. SMI}, June 2006.

\bibitem{meyer2003discrete}
Mark Meyer, Mathieu Desbrun, Peter Schr{\"o}der, and Alan~H Barr.
\newblock Discrete differential-geometry operators for triangulated
  2-manifolds.
\newblock In {\em Visualization and Mathematics III}, pages 35--57. Springer,
  2003.

\bibitem{milnor.tori}
John Milnor.
\newblock Eigenvalues of the {L}aplace operator on certain manifolds.
\newblock {\em PNAS}, 51(4):542, 1964.

\bibitem{nogneng2017informative}
Dorian Nogneng and Maks Ovsjanikov.
\newblock Informative descriptor preservation via commutativity for shape
  matching.
\newblock {\em Computer Graphics Forum}, 36(2):259--267, 2017.

\bibitem{ovsjanikov2012functional}
Maks Ovsjanikov, Mirela Ben-Chen, Justin Solomon, Adrian Butscher, and Leonidas
  Guibas.
\newblock {F}unctional {M}aps: {A} {F}lexible {R}epresentation of {M}aps
  {B}etween {S}hapes.
\newblock {\em TOG}, 31(4):30, 2012.

\bibitem{ovsjanikov2017computing}
Maks Ovsjanikov, Etienne Corman, Michael Bronstein, Emanuele Rodol\`{a}, Mirela
  Ben-Chen, Leonidas Guibas, Frederic Chazal, and Alex Bronstein.
\newblock Computing and processing correspondences with functional maps.
\newblock In {\em SIGGRAPH Courses}, pages 5:1--5:62, 2017.

\bibitem{panine2016towards}
Mikhail Panine and Achim Kempf.
\newblock Towards spectral geometric methods for euclidean quantum gravity.
\newblock {\em Physical Review D}, 93(8):084033, 2016.

\bibitem{reuter05}
Martin Reuter, Franz-Erich Wolter, and Niklas Peinecke.
\newblock Laplace-spectra as fingerprints for shape matching.
\newblock In {\em Proc. SPM}, SPM '05, pages 101--106, New York, NY, USA, 2005.
  ACM.

\bibitem{reuter06}
Martin Reuter, Franz-Erich Wolter, and Niklas Peinecke.
\newblock Laplace-beltrami spectra as 'shape-dna' of surfaces and solids.
\newblock {\em Computer Aided Design}, 38(4):342--366, Apr. 2006.

\bibitem{schonsheck2018nonisometric}
Stefan~C Schonsheck, Michael~M Bronstein, and Rongjie Lai.
\newblock Nonisometric surface registration via conformal laplace-beltrami
  basis pursuit.
\newblock {\em arXiv:1809.07399}, 2018.

\bibitem{sorkine2004least}
Olga Sorkine and Daniel Cohen-Or.
\newblock Least-squares meshes.
\newblock In {\em Proc. Shape Modeling Applications, 2004. Proceedings}, pages
  191--199, 2004.

\bibitem{tam2013registration}
Gary~KL Tam, Zhi-Quan Cheng, Yu-Kun Lai, Frank~C Langbein, Yonghuai Liu, David
  Marshall, Ralph~R Martin, Xian-Fang Sun, and Paul~L Rosin.
\newblock Registration of 3d point clouds and meshes: a survey from rigid to
  nonrigid.
\newblock {\em IEEE Trans. Visualization and Computer Graphics},
  19(7):1199--1217, 2013.

\bibitem{van2011survey}
Oliver Van~Kaick, Hao Zhang, Ghassan Hamarneh, and Daniel Cohen-Or.
\newblock A survey on shape correspondence.
\newblock In {\em Computer Graphics Forum}, volume~30, pages 1681--1707, 2011.

\bibitem{weyl}
Hermann Weyl.
\newblock {\"U}ber die asymptotische {V}erteilung der {E}igenwerte.
\newblock {\em Nachrichten von der Gesellschaft der Wissenschaften zu
  G{\"o}ttingen, Mathematisch-Physikalische Klasse}, pages 110--117, 1911.

\bibitem{zelditch1998revolution}
Steve Zelditch.
\newblock The inverse spectral problem for surfaces of revolution.
\newblock {\em J. Diff. Geom.}, 49(2):207--264, 1998.

\bibitem{zelditch2000spectral}
Steve Zelditch.
\newblock Spectral determination of analytic bi-axisymmetric plane domains.
\newblock {\em Geometric \& Functional Analysis GAFA}, 10(3):628--677, 2000.

\bibitem{zelditch2009inverse}
Steve Zelditch.
\newblock Inverse {S}pectral {P}roblem for {A}nalytic {D}omains, {II}:
  $\mathbb{Z}_{2}$-symmetric domains.
\newblock {\em Annals of {M}athematics}, 170(1):205--269, 2009.

\end{thebibliography}
}

\end{document}